\documentclass[fleqn,usenatbib,useAMS]{mnras}

\usepackage{graphicx}	
\usepackage{amsmath}	
\usepackage{amssymb}	
\usepackage{multicol}        
\usepackage{bm}		
\usepackage{pdflscape}	
\usepackage{xcolor}

\usepackage[T1]{fontenc}
\usepackage{ae,aecompl}

\usepackage{newtxtext,newtxmath}

\title[Power spectrum of surface-brightness anomalies]{Probing sub-galactic mass structure with the power spectrum \\ of surface-brightness anomalies in high-resolution observations \\ of galaxy-galaxy strong gravitational lenses. I.~ Power-spectrum measurement and feasibility study}

\author[D. Bayer et al.]{D. Bayer,$^{1,}$$^{2,}$$^{3}$\thanks{Contact e-mail: \href{mailto:dbayer@swin.edu.au}{dbayer@swin.edu.au}}
L. V. E. Koopmans,$^{3}$
J. P. McKean,$^{3,}$$^{4}$
S. Vegetti,$^{5}$ 
T. Treu,$^{6}$
\newauthor C. D. Fassnacht$^{7}$ and  K. Glazebrook$^{1,}$$^{2}$  
\\
$^{1}$Centre for Astrophysics \& Supercomputing, Swinburne University of Technology, Hawthorn, VIC 3122, Australia\\
$^{2}$ARC Centre of Excellence for All Sky Astrophysics in 3 Dimensions (ASTRO 3D), Australia\\
$^{3}$Kapteyn Astronomical Institute, University of Groningen, PO Box 800, 9700 AV Groningen, the Netherlands\\
$^{4}$ASTRON, Netherlands Institute for Radio Astronomy, Postbus 2, NL-7990 AA, Dwingeloo, the Netherlands\\
$^{5}$Max Planck Institute for Astrophysics, Karl-Schwarzschild-Strasse 1, D-85740 Garching, Germany\\
$^{6}$Department of Physics and Astronomy, UCLA, 430 Portola Plaza, Los Angeles, CA 90095-1547, USA\\
$^{7}$Department of Physics and Astronomy, University of California, Davis, 1 Shields Ave. Davis, CA 95616, USA}

\date{Accepted XXX. Received YYY; in original form ZZZ}

\pubyear{2023}

\begin{document}
\label{firstpage}
\pagerange{\pageref{firstpage}--\pageref{lastpage}}
\maketitle

\begin{abstract}
While the direct detection of the dark-matter particle remains very challenging, the nature of dark matter could be possibly constrained by comparing the observed abundance and properties of small-scale sub-galactic mass structures with predictions from the phenomenological dark-matter models, such as cold, warm or hot dark matter. Galaxy-galaxy strong gravitational lensing provides a unique opportunity to search for tiny surface-brightness anomalies in the extended lensed images (i.e. Einstein rings or gravitational arcs), induced by possible small-scale mass structures in the foreground lens galaxy. In this paper, the first in a series, we introduce and test a methodology to measure the power spectrum of such surface-brightness anomalies from high-resolution \textit{Hubble Space Telescope} (HST) imaging. In particular, we focus on the observational aspects of this statistical approach, such as the most suitable observational strategy and sample selection, the choice of modelling techniques and the noise correction. We test the feasibility of the power-spectrum measurement by applying it to a sample of galaxy-galaxy strong gravitational lens systems from the Sloan Lens ACS Survey, with the most extended, bright, high-signal-to-noise-ratio lensed images, observed in the rest frame ultraviolet. In the companion paper, we present the methodology to relate the measured power spectrum to the statistical properties of the underlying small-scale mass structures in the lens galaxy and infer the first observational constraints on the sub-galactic matter power spectrum in a massive elliptical (lens) galaxy.
\end{abstract}





\section{Introduction}

Over the last forty years, studies of the spatial mass distribution inside galaxies have provided valuable insights into the complex processes of galaxy formation and evolution. Most importantly, the internal mass density profiles of spiral galaxies inferred from their kinematics \citep{Bosma_1978PhDThesis, Rubin} have led to the hypothesis of a hitherto unknown dominant non-baryonic matter component, referred to as \emph{dark matter}, which nowadays constitutes a crucial pillar of the concordance dark-energy-plus-cold-dark-matter ($\Lambda$CDM) cosmological model. According to this model and the associated hierarchical structure-formation scenario, the early gravitational collapse of dark matter into haloes has created the potential wells necessary for the baryonic gas to cool and condense, finally leading to the formation of the observable galaxies \citep{Blumenthal1984,1991WhiteFrenk,Gao}.

Despite this essential role of dark matter in the cosmological structure-formation process, its nature and properties remain unknown. The standard cold-dark-matter (CDM) paradigm is still challenged by various alternative models, such as for example warm dark matter \citep[WDM, see e.g.][]{WDMsim, Lovell2014} or self-interacting dark matter \citep[SIDM, see e.g.][]{SelfinteractingDM, Tulin_SelfinteractingDM}. These have been proposed in an attempt to explain the striking discrepancy between the number of dwarf satellite galaxies observed in the Local Group and the corresponding predictions from $\Lambda$CDM-based simulations \citep[i.e. the Missing Satellites Problem, see e.g.][]{Klypin,Moore,Diemand_Kuhlen_Madau_2007, Local_Group_Satellites,DES_MWSatellites, MSP_3D,  MSP_LMG}. 

In general, the abundance of such sub-galactic mass structures is determined by the free-streaming length of dark matter in the early Universe. This, in turn, depends on the microscopic properties of the dark-matter particles, such as the particle mass or the strength of the particle-particle interactions. Whereas the standard CDM model predicts galaxy-size haloes to be inhabited by an abundant population of mass structures, the alternative models with less massive or self-interacting dark-matter particles significantly suppress the formation of sub-galactic mass structures, especially in the low-mass regime below $\sim10^8 M_\odot$ \citep[see, for example,][]{Bullock2017_Review}. Hence, while the direct detection of the dark-matter particle remains challenging \citep[see e.g.][]{Bertone_DM}, its properties could be constrained based on the observed abundance and properties of low-mass structures in a representative sample of galaxies.

However, detecting such low-mass sub-galactic structures beyond the Local Group is a demanding undertaking. They are generally thought to be dark-matter dominated or even purely dark (i.e. completely devoid of stars) and, thus, intrinsically invisible. Even if massive enough to form stars, they might be too faint to be observed directly at cosmological distances with the currently available instruments. For this reason, constraints on the properties of sub-galactic mass structures at cosmological distances have been so far inferred mainly from high-resolution observations and modelling of galaxy-scale strong gravitational lenses \citep[e.g.][]{Mao, Metcalf_Madau, Dalal_Kochanek, Vegetti_2010_dwarf, Vegetti2010_detection, Nierenberg2014_detection, Birrer_2keV, GilmanDM, Gilman_WDM_chills, Ritondale, Hsueh2020WDM}. 

In particular, the phenomenon of galaxy-galaxy strong gravitational lensing makes it possible to detect mass structures in galaxies that, fortuitously, happen to lie along the same line-of-sight and act as a strong gravitational lens on another galaxy located at a larger distance. Mass structures in the lens galaxy (and possible line-of-sight haloes) induce perturbations to the otherwise smooth lensing potential. These, in turn, perturb the deflection angles of light rays crossing the lens plane in proximity to the mass structures. Thus, even if the structures were purely dark, their gravitational signatures might be observable in the form of the resulting anomalies in the surface-brightness distribution of the extended lensed images (i.e. Einstein ring or gravitational arcs), measured with respect to the best-fitting smooth-lens model \citep{Blandford, Koopmans2005, Rau2013}. One of the most successful methods utilizing this effect to search for individual mass structures in (massive elliptical) lens galaxies is the \emph{gravitational-imaging technique} \citep{Koopmans2005, VegettiKoopmans2009}. The application of this technique to deep \textit{Hubble Space Telescope} (HST) imaging has so far resulted in a detection of two dark-matter subhaloes with the mass of $3.5 \times 10^9 M_\odot$ and $1.9 \times 10^8 M_\odot$ at the redshift $z = 0.2$ and $z = 0.881$, respectively, with the latter one being the smallest and most distant galactic substructure discovered up to now beyond the local Universe \citep{Vegetti_DR_substructure,Vegetti_Nature}. 

In order to investigate less massive sub-galactic mass structures, numerously predicted in $\Lambda$CDM-based cosmological simulations, \cite{Sander_thesis}, \cite{Hezaveh_PS}, \cite{Diaz_Rivero_2018} and \cite{Saikat} proposed a complementary statistical approach. Instead of individual massive subhaloes in the lens galaxy, represented by localised potential corrections, the statistical approach models the entire population of small-scale sub-galactic mass structures as Gaussian-random-field (GRF) potential perturbations superposed on the best-fitting smoothly-varying lensing potential. In the framework of the theoretical formalism proposed by \cite{Saikat}, both the potential perturbations and the collectively-induced surface-brightness anomalies in the lensed images are quantified in terms of their power spectra and related to each other. Successful tests of this approach on mock lensed images, presented by \cite{Saikat} and \cite{Saikat_thesis}, suggest that it might be possible to infer observational constraints on the power spectrum of small-scale mass structures in a (massive elliptical) lens galaxy from the power spectrum of the resulting surface-brightness anomalies in the extended lensed images of the background source galaxy.

This paper is the first in a series of papers aimed at investigating the potential and feasibility of applying this power-spectrum approach to real observational data. The goal of the present paper is to introduce and test the methodology to reliably extract the power spectrum of surface-brightness anomalies from high-resolution HST-imaging of galaxy-galaxy strong gravitational lenses. In particular, we focus on the observational aspects of the power-spectrum measurement, such as the most suitable observational strategy, the sample selection, the choice of modelling techniques and the noise correction. In the companion paper (Paper II), we extend the methodology in order to relate the measured power spectrum of surface-brightness anomalies in the lensed images to the statistical properties of the underlying small-scale mass structure in the lens galaxy and infer the first observational constraints on the matter power spectrum in a massive elliptical (lens) galaxy. Future research will apply this approach to a larger sample of lens systems and compare the results with predictions from hydrodynamical simulations, which might eventually allow us to distinguish between the alternative dark-matter models. 

The paper is structured as follows. In Section~\ref{Section:SB_anomalies}, we formalise the concept of surface-brightness anomalies in extended lensed images of a galaxy-galaxy strong gravitational lens system. Section~\ref{Section:Observations} moves on to describe our observational strategy, sample selection and the imaging data. In Section~\ref{Section:Methodology_UV_sample}, we present our methodology to measure the power spectrum of surface-brightness anomalies caused by small-scale mass structures in the lens galaxy. Section~\ref{Section:Performance_test} demonstrates the feasibility of our approach in recovering mock surface-brightness anomalies from simulated lensed images mimicking real observations. Finally, Section~\ref{Section:Conclusions_UV_sample} provides conclusions and implications for further work.

For a consistent comparison of the inferred smooth lens models with earlier studies by \cite{Vegetti2014}, throughout this paper we assume the following cosmology: $H_0 = 73\ \mathrm{km \  s^{-1}Mpc^{-1}}$, $\Omega_M = 0.25$ and $\Omega_\Lambda = 0.75$. Given this cosmology, 1~arcsec corresponds to $\sim 4$ kpc at the redshift of the studied lens galaxies ($z_{L} \sim 0.2-0.4$).

\section{Surface-brightness anomalies in extended lensed images}
\label{Section:SB_anomalies}

In this section, we first discuss the concept of the hypothetical surface-brightness anomalies that would emerge in the extended lensed images of a background source galaxy as a result of small-scale density fluctuations in the foreground lens galaxy. Subsequently, we elaborate on the observational and modelling challenges that need to be circumvented in order to accurately measure such anomalies.

Let us consider a galaxy-galaxy strong gravitational lens system with extended lensed images described by the surface-brightness distribution $I(\textbf{\textit{x}})$ as a function of the position $\textbf{\textit{x}}$ in the lens plane. The spatial configuration of the lens system is parametrized by the angular diameter distances from the observer to the foreground lens galaxy $D_{d}$, from the observer to the background source galaxy $D_{s}$ and from the lens to the source galaxy $D_{ds}$. 
Following the convention of strong gravitational lensing, we express the surface mass density $\Sigma(\textbf{\textit{x}})$ of the lens galaxy (including the possible line-of-sight haloes) in units of the critical surface mass density:
\begin{equation}
\Sigma_{\rm{cr}} = \frac{c^{2}}{4 \pi G} \frac{D_{s}}{D_{d} D_{ds}}
\label{critical density}
\end{equation} to obtain the commonly used (dimensionless) convergence:
\begin{equation}
\kappa(\textbf{\textit{x}})=\Sigma( \textbf{\textit{x}}) / \Sigma_{\rm{cr}}.
\label{eq:kappa}
\end{equation}
Furthermore, we define the lensing potential, i.e. the gravitational potential of the lens galaxy projected along the line of sight: 
\begin{equation}
\psi(\textbf{\textit{x}})=\frac{1}{\pi}\int_{\mathbb{R}^2}\bm{\textbf{\textit{dx}}^{\prime}}  \kappa (\textbf{\textit{x}}^{\prime})\  \rm{ln}\mid \textbf{\textit{x}}-\textbf{\textit{x}}^{\prime}\mid,
\label{lensing_potential}
\end{equation}
which is related to the convergence by the Poisson's equation:
\begin{equation}
\nabla^2 \psi( \textbf{\textit{x}}) = 2 \kappa( \textbf{\textit{x}}). 
\label{Poisson}
\end{equation}
The associated (scaled) deflection-angle field:
\begin{equation}
\bm{\alpha} (\textbf{\textit{x}})= \nabla \psi(\textbf{\textit{x}})
\label{deflection_angle}
\end{equation}
determines a mapping between the positions $\textbf{\textit{x}}$ and $\textbf{\textit{y}}$ in the lens- and the source plane, respectively, which is encapsulated in the lens equation:
\begin{equation}
\textbf{\textit{y}}(\textbf{\textit{x}}) = \textbf{\textit{x}}- \bm{\alpha}(\textbf{\textit{x}}).
\label{lens_equation}
\end{equation}
This mapping together with the principle of surface-brightness conservation in strong gravitational lensing:
\begin{equation}
I(\textbf{\textit{x}}) = S(\textbf{\textit{y}}(\textbf{\textit{x}}))
\label{SB_conservation}
\end{equation}
builds the foundation for numerical grid-based smooth-lens-modelling codes \citep[e.g. the adaptive grid-based Bayesian lens-modelling code by][used in this work]{VegettiKoopmans2009} which allow one to simultaneously reconstruct the best-fitting smooth (parametric) lensing potential $\psi_{M}(\textbf{\textit{x}})$ and the unlensed intrinsic surface-brightness distribution of the source galaxy $S(\textbf{\textit{y}})$. 

A discrepancy between the surface-brightness distribution of the observed lensed images $I(\textbf{\textit{x}})$ and the prediction from the best-fitting smooth-lens model $I_M(\textbf{\textit{x}})$ might point towards the presence of mass structure in the lens galaxy. As can be seen from equations \ref{deflection_angle} and \ref{lens_equation}, a deviation of the true lensing potential $\psi(\textbf{\textit{x}})$ from the best-fitting smooth lensing potential $\psi_{M}(\textbf{\textit{x}})$ modifies the mapping between the lens- and the source plane. This, in turn, results in a surface-brightness change $\delta I(\textbf{\textit{x}})$, such that:
\begin{equation}
\delta I(\textbf{\textit{x}}) = I(\textbf{\textit{x}}) - I_M(\textbf{\textit{x}}) = S\big( \textbf{\textit{x}} - \nabla \psi(\textbf{\textit{x}})\big) - S\big(\textbf{\textit{x}} - \nabla \psi_M(\textbf{\textit{x}})\big)
\label{SB}
\end{equation}
\citep{Blandford, Koopmans2005}.
In what follows, we refer to $\delta I(\textbf{\textit{x}})$ as \emph{surface-brightness anomalies}. 

In reality, the extraction of such surface-brightness anomalies from the imaging of real lens systems is complicated by the following issues.

\begin{itemize}

\item First, observational effects make it impossible to measure the true surface-brightness distribution of the lensed images $I(\textbf{\textit{x}})$. Instead, the lensed images are blurred by the convolution with the point-spread function (PSF) and the pixellation of the imaging. Moreover, they are affected by the presence of the observational noise and subject to data processing, for example drizzling of the raw data to obtain the final science image \citep{Gonzaga2012}. 

\item Second, the reconstructed unlensed surface-brightness distribution of the source galaxy cannot be assumed to perfectly represent the reality. This is due to a degeneracy between the perturbative lensing effect of mass structure in the lens galaxy and the intrinsic surface-brightness fluctuations in the source galaxy. To mitigate this problem, the adaptive grid-based Bayesian smooth-lens-modelling code by \cite {VegettiKoopmans2009}, used in this work, applies a regularisation of the source reconstruction by penalizing solutions with overly strong surface-brightness fluctuations in the source galaxy, as discussed by \cite{Warren_Dye} and \cite{Koopmans2005}. However, the imposed level of regularisation itself is optimized for in the lens-modelling procedure, which might either suppress or enhance the true surface-brightness fluctuations in the source as a result of an over- or underregularised source reconstruction, respectively. In other words, the reconstructed smooth-lens model might potentially "absorb" the effect of mass structure into spurious source structure or vice versa -- the surface-brightness anomalies due to mass structure might be artificially enhanced if the source reconstruction is overregularised (i.e. too smooth). 

\item Third, the effect of strong gravitational lensing is sensitive to the total mass present in the line-of-sight along which the gravitational lens is observed. Thus, the investigated surface-brightness anomalies might arise not only from mass structure in the lens galaxy, but also from possible line-of-sight haloes \citep{Li2016,Despali_LOS}. While a detection along the line of sight is valuable for its own sake, it makes the interpretation of the results more complicated.

\end{itemize}

Hence, one of the main challenges in our approach is to reliably extract the true surface-brightness anomalies $\delta I(\textbf{\textit{x}})$ (as defined in equation \ref{SB}) from the smooth-lens-model residuals, taking into consideration all the effects discussed above.

\section{Observational strategy and data}
\label{Section:Observations}

We perform this pilot study based on our HST/WFC3/F390W-observations \citep[Program 12898,][]{UV_HST_proposal} of 10 lens systems with highly-structured star-forming lensed galaxies selected from the SLACS Survey \citep{Bolton2008}. In this section, we first motivate the choice of the ultra-violet band and discuss our selection criteria for the SLACS sub-sample. Subsequently, we elaborate on the undersampling problem, the dithering strategy and the data reduction.

\subsection{Selection of the observational filter}
\label{Section:WhyUV}

The level of surface-brightness anomalies caused by the presence of mass structures in the foreground lens galaxy depends not only on the substructure mass, but also on the gradient (i.e. level of variations) in the intrinsic surface-brightness distribution of the lensed source galaxy itself. More specifically, if the population of mass structures in the lens galaxy is represented by a potential-perturbation field $\delta\psi(\textbf{\textit{x}})$ and the intrinsic surface-brightness distribution of the source galaxy is described by $S(\textbf{\textit{y}})$, then the level of the resulting surface-brightness anomalies $\delta I(\textbf{\textit{x}})$ can be computed as the inner product of the respective gradient fields: 
\begin{equation}
\delta I(\textbf{\textit{x}}) = -\nabla S(\textbf{\textit{y}}(\textbf{\textit{x}}))  \cdot \nabla\delta\psi(\textbf{\textit{x}}) 
\label{eq:SB_change}
\end{equation} 
\citep{Blandford, Koopmans2005}.
\noindent Hence, the surface-brightness anomalies caused by a given population of mass structures in the lens galaxy can be enhanced by a high level of variations in the intrinsic surface brightness of the source galaxy. 

This motivates our choice of the ultraviolet band and the selection of lens systems with highly-structured star-forming lensed galaxies. While the optical and infrared bands, used in earlier gravitational-imaging studies of SLACS lenses \citep[e.g.][]{Vegetti2014}, capture mostly the smooth old stellar populations, the compact star-forming regions prominent in the ultraviolet band are expected to enhance the surface-brightness gradients in the lensed galaxies, allowing us to improve the sensitivity of our approach to low-mass structures in the lens galaxies \citep[see also ][]{Ritondale}. We choose to carry out our observations using the Wide Field Camera~3 (WFC3) onboard the HST, which offers the highest resolution and signal-to-noise ratio currently available in the rest-frame ultraviolet \citep[HST/WFC3/F390W, Program 12898,][]{UV_HST_proposal}.

\subsection{Sample selection}
\label{Section:SLACS_subsample}

For our HST/WFC3/F390W-observations, we select a sub-sample of ten lens systems from the SLACS Survey \citep{Bolton2008}, which is the so-far largest homogeneous sample of galaxy-galaxy strong gravitational lenses, comprising more than a hundred lens systems. Most of them consist of a massive early-type lens galaxy aligned with a blue star-forming source galaxy, in compliance with our observational strategy. The already existing multi-colour HST-imaging data as well as the extensive lens-modelling and kinematic studies \citep[e.g.][]{Auger2009, Vegetti2014} and the availability of the spectroscopic redshifts for both the lens and the source galaxies make the SLACS lenses an excellent choice to test our methodology. 

We intentionally exclude all SLACS systems with late-type lens galaxies to avoid possible degeneracies between the star formation or dust extinction in the lens galaxy and in the lensed images. For all remaining lens systems, we obtain the SDSS spectra and determine the [OII]-flux of the source galaxies, which is assumed to be a good proxy for their star-formation rate. Furthermore, we retrieve the already existing one-orbit HST/ACS/F814W-observations and calculate the average surface brightness within the most compact area of the lensed images containing half of the total lensed flux. This can be regarded as the \emph{effective surface brightness} of the lensed images. A good correlation between the measured [OII]-flux and the surface brightness of the lensed images in F814W indicates that the latter is related to the young stellar populations in the source galaxy (but is more diffuse). Thus, by selecting SLACS lens systems with the brightest lensed images in the F814W-filter we at the same time select lens systems with highly star-forming source galaxies, in accordance with our observational strategy.

Our final target list comprising 10 SLACS lens systems with the brightest and most extended lensed images is presented in Table~\ref{tab:full_sample} together with the basic astrometric and spectroscopic properties: the location on the sky, the spectroscopic redshifts of the lens and the source galaxies, and the stellar velocity dispersion in the lens galaxy \citep{Auger2009}. To improve the quality of the new F390W-imaging data in comparison to the already existing observations in the F814W-filter, we require an average signal-to-noise ratio of 10 in the pixels covering the lensed images and calculate the observing time accordingly. We quote the number of HST-orbits devoted to the observations of each lens system in the last column of Table~\ref{tab:full_sample}. 

\begin{table*}
\centering
\caption{Astrometric and spectroscopic properties of the selected SLACS sub-sample observed with the Hubble Space Telescope in the rest-frame ultra-violet (HST/WFC3/F390W): the target names, location on the sky, spectroscopic redshifts of the lens $z_{l}$ and the source galaxy $z_{s}$, the velocity dispersion of the lens galaxy $\sigma_{l}$ \citep[from][]{Auger2009}, and the number of HST-orbits devoted to the observations of each lens system. For the purpose of this study, we perform the lens modelling for only four most suitable lens systems from this sample -- SDSS J0252+0039, SDSS J0737+3216, SDSS J1430+4105 and SDSS J1627--0053 -- characterised by a relatively simple geometry and showing no substantial galaxy-core residuals in the galaxy-subtracted images.}
    \begin{tabular}{lcccccc}
\hline
Target lens galaxy & Right Ascension & Declination & $z_{l}$ & $z_{s}$ & $\sigma_{l} \rm[km/s]$ & Orbits\\
\hline

SDSS J0252+0039 & 02 52 45.21 & +00 39 58.40 &0.280 & 0.982 & 164 $\pm 12$& 2 \\
SDSS J0737+3216 & 07 37 28.45 & +32 16 18.60 &0.322 & 0.581 & 338 $\pm 16$& 2\\
SDSS J0903+4116 & 09 03 15.19 & +41 16 09.10 &0.430 & 1.065 & 223 $\pm 27$& 6 \\
SDSS J0912+0029 & 09 12 05.31 & +00 29 01.20 &0.164 & 0.324 & 326 $\pm 12$& 5\\
SDSS J0956+5100 & 09 56 29.78 & +51 00 06.60 &0.241 & 0.470 & 334 $\pm 15$&2 \\
SDSS J0959+0410 & 09 59 44.07 & +04 10 17.00  &0.126 & 0.535 & 197 $\pm 13$ &1\\
SDSS J1430+4105 & 14 30 04.10 & +41 05 57.10 &0.285 & 0.575 & 322 $\pm 32$&2  \\
SDSS J1627-- 0053 & 16 27 46.45 & -- 00 53 57.60 &0.208 & 0.524 & 290 $\pm 14$& 3\\
SDSS J1630+4520 & 16 30 28.16 & +45 20 36.30 &0.248 & 0.793 & 276 $\pm 16$& 5\\
SDSS J2341+0000 & 23 41 11.57 & +00 00 18.70&0.186 & 0.807 & 207 $\pm13$  &2 \\
\hline
\end{tabular}
\label{tab:full_sample}
\end{table*}

\subsection{Observations and data reduction}
\label{Section:Data reduction}

Our HST/WFC3/F390W-observations were performed between
January 26 and September 16, 2013 \citep[Program 12898,][]{UV_HST_proposal}. Raw HST-images are generally known to be undersampled, which means that the (blurred) point sources are not covered by enough pixels to precisely sample the point-spread function (PSF). Whereas a well-sampled image would have at least two pixels across the full width at half maximum ($\mathrm{FWHM}$) of the PSF (Nyquist limit), the pixel width of the WFC3 ($0.04$ arcsec on a side) is relatively large in comparison to the FWHM of the WFC3/UVIS optical performance at 390 nm ($0.07$ arcsec). In order to alleviate this undersampling problem and more optimally benefit from the superb resolution of the HST-optics, we apply the standard \emph{dithering} strategy, i.e. we obtain multiple dithered exposures of each target object by slightly shifting the telescope pointing every time the next exposure is taken. Besides the improvement in the PSF-sampling, this dithering technique makes it possible to compensate for cosmic rays, possible dead pixels or columns, and the flat-field effects.

We retrieve the dithered exposures from the MAST archive\footnote{http://archive.stsci.edu/hst/search.php} in the form of pipeline-preprocessed flat-field calibrated FITS files (flt.fits files) and make use of the Variable-Pixel Linear Reconstruction algorithm \citep{Drizzle}, informally known as \textit{Drizzle}, to combine them into the final science images of our sample. We perform the drizzling in an automatic way by means of the \textsc{astrodrizzle} task from the \textsc{drizzlepac} package \citep{Gonzaga2012} in the default configuration. In Section~\ref{Section:Noise_correlation_UV}, we investigate the impact of different drizzling settings on our final results in comparison to this reference configuration.

\begin{figure*}
\centering
 \includegraphics[scale=1.0]{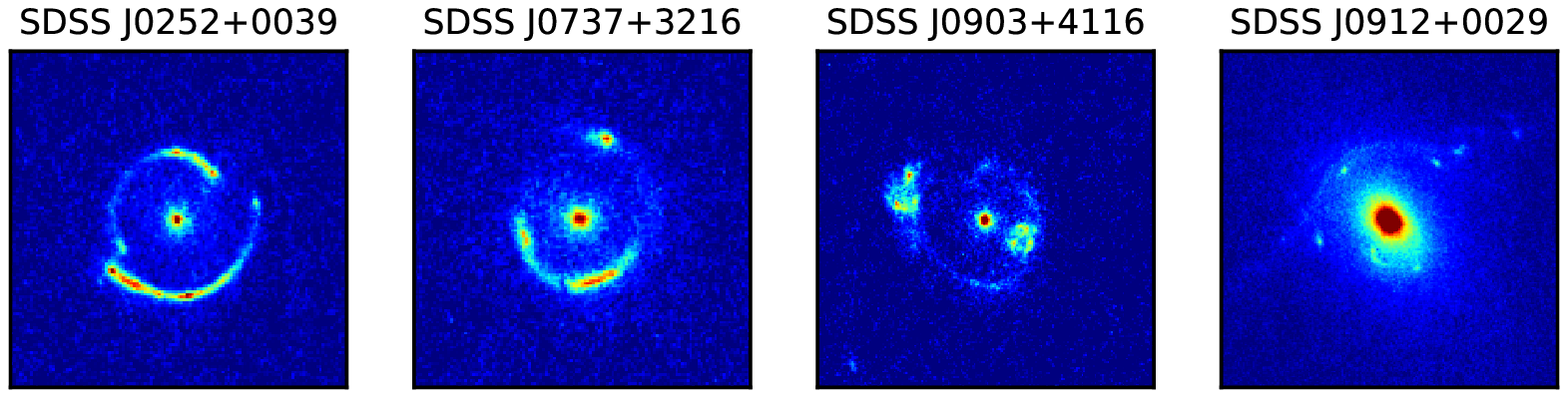}
 \includegraphics[scale=1.0]{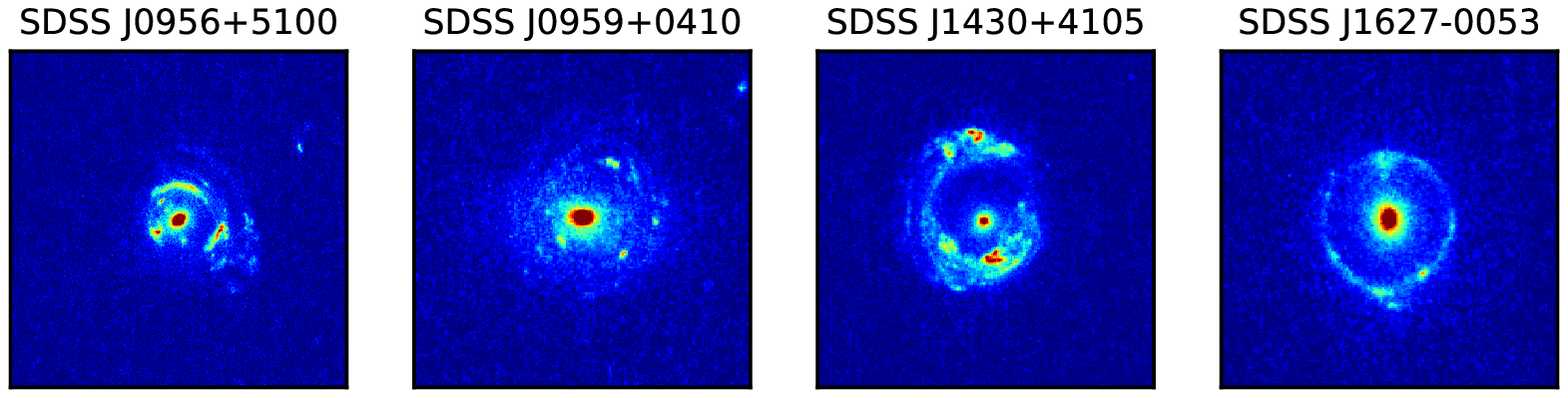}
 \includegraphics[scale=1.0]{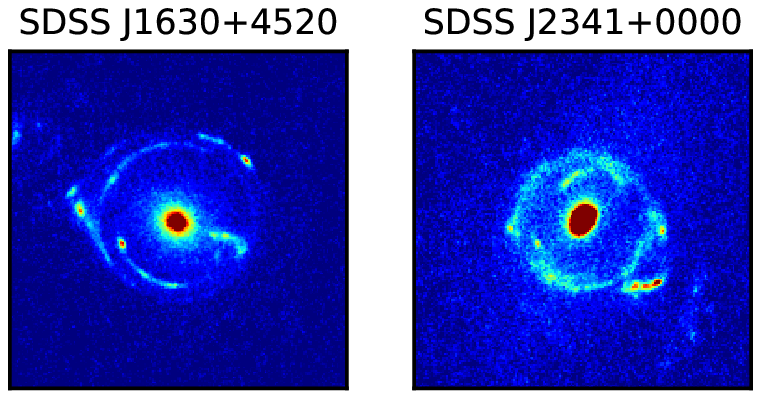}
 \caption{HST/WFC3-imaging of the selected SLACS sub-sample in the rest-frame ultra-violet (F390W). Each of these images depicts a massive elliptical galaxy acting as a strong gravitational lens on a star-forming background source galaxy. The source galaxies appear lensed into high-signal-to-noise-ratio gravitational arcs or (in some cases) an almost complete Einstein ring. The image cutouts are centred on the brightest pixel of the lens galaxy and have the side length approximately equal to four Einstein radii of the respective lens system.}
\label{fig:sampleUV}
\end{figure*}

\begin{figure*}
 \centering
 \includegraphics[scale=0.65]{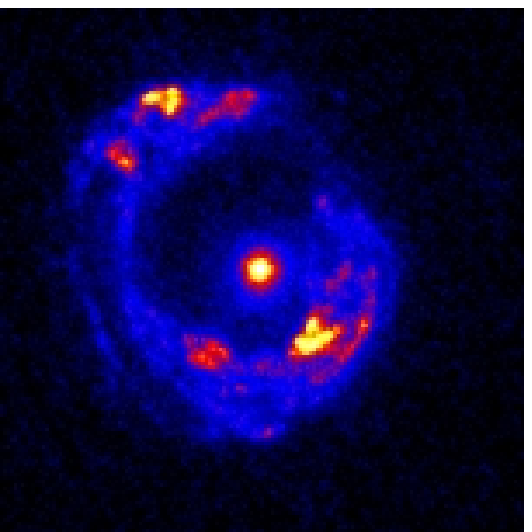}
 \includegraphics[scale=0.65]{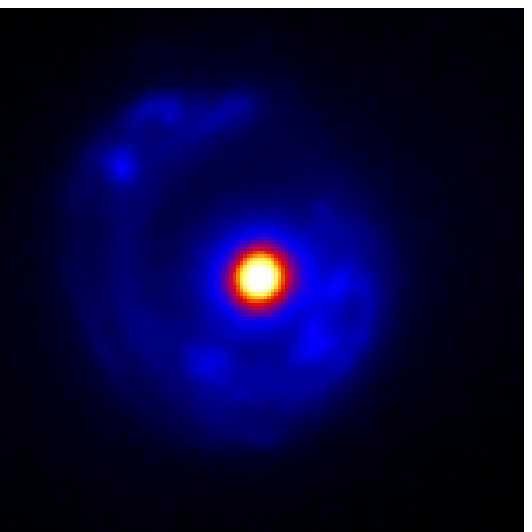}
 \includegraphics[scale=0.65]{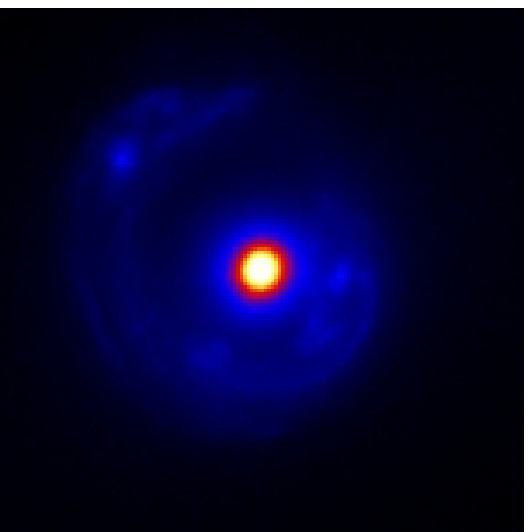}
 \includegraphics[scale=0.65]{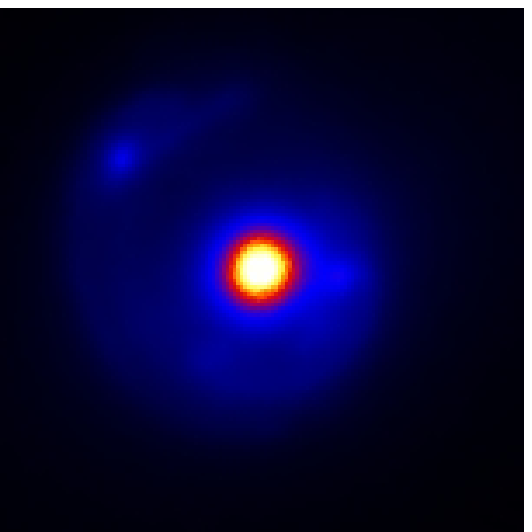}
 \includegraphics[scale=0.65]{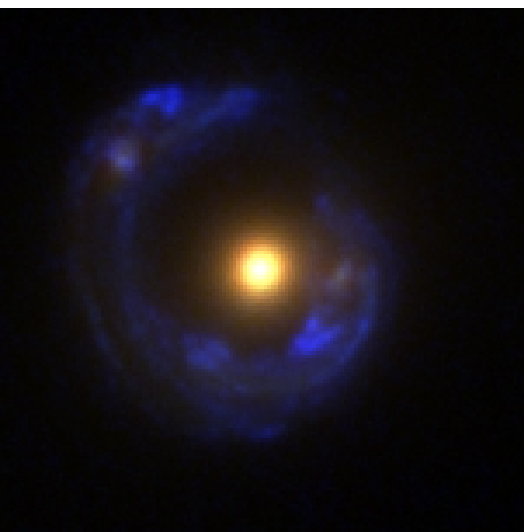}

 \caption{A high-resolution multi-band view of the Einstein ring in the strong gravitational lens system SDSS J1430+4105. \textit{From left to right}: our new HST/WFC3-observations in F390W, the archival HST-imaging in F606W, F814W and F160W, and the colour-composite image combining the multi-band photometry in F390W, F606W and F814W obtained using the \textsc{stiff} software. All images are oriented with north pointing up and east pointing left.}
 \label{fig:multicolorJ1430}
\end{figure*}

Next, for each lens system, we generate an image cutout centred on the brightest pixel of the lens galaxy, with the side length approximately equal to four Einstein radii, as presented in Fig.~\ref{fig:sampleUV}. Additionally, for the purpose of illustration, Fig.~\ref{fig:multicolorJ1430} depicts the obtained F390W-imaging for the lens system SDSS J1430+4105 in comparison to the archival multi-band observations in F606W, F814W and F160W. The colour-composite image, created using the \textsc{stiff}\footnote{https://www.astromatic.net/software/stiff} software, combines the F390W-imaging with the archival F814W and F160W-observations.

Finally, we use the  \textsc{tinytim}\footnote{http://www.stsci.edu/hst/observatory/focus/TinyTim} \citep{TinyTim_Paper} software to model the PSF of the HST/WFC3/F390W-optics in the central pixel of each image cutout. For simplicity, we assume that the PSF in all the other pixels of the considered image cutouts does not deviate significantly from the central pixel. In our PSF-models, we account for the different spectral types of the lens galaxies, which we estimate based on the magnitudes in F555W, F614W, F814W and F160W as inferred by \cite{Auger2009}. 

\section{Methodology}
\label{Section:Methodology_UV_sample}

In this section, we present our methodology to measure the power spectrum of surface-brightness anomalies in high-resolution HST-observations of galaxy-galaxy strong gravitational lens systems. 

\subsection{Analysis synopsis}
\label{Section:Analysis synopsis}

Our procedure consists of the steps outlined below.

\begin{enumerate}

\item Modelling and subtraction of the lens-galaxy light by means of \textsc{galfit} \citep{GALFIT_paper} or, alternatively, the \textsc{b-spline} algorithm \citep{Bolton2006}, see Section \ref{Section:Lens-galaxy subtraction}; \vspace{0.1cm}

\item Smooth lens modelling using the adaptive and grid-based Bayesian lens-modelling technique by \cite{VegettiKoopmans2009}, see Section \ref{Section:Lens_Model_UV_sample};\vspace{0.1cm}

\item Statistical quantification of the residual surface-brightness fluctuations in the lensed images in terms of the azimuthally-averaged power spectrum, see Section \ref{Section:PS}; \vspace{0.1cm}

\item Estimation of the noise power spectrum based on blank-sky fields (modified to account for the additional Poisson noise present in the imaging of a lens system), see Section \ref{Section:Noise_correction};\vspace{0.1cm}

\item Noise-bias correction to reveal the power spectrum of surface-brightness anomalies, see Section \ref{Section:PS_SB_anomalies}.\vspace{0.1cm}

\end{enumerate}
In the following sections, we discuss the individual steps in more detail and illustrate them with examples from the analysis of our SLACS sub-sample.

\subsection{Lens-galaxy subtraction}
\label{Section:Lens-galaxy subtraction}

Before proceeding with the lens modelling, here we discuss how to reliably estimate and correct for the flux contribution from the lens galaxy in the pixels overlapping with the lensed images. In a typical SLACS lens system, the Einstein radius ($\sim$1 arcsec) is comparable to half the effective radius of the (massive elliptical) lens galaxy \citep{Auger2009}. This means that the lensed images in our sample are projected on the very inner region ($\sim5-10$ kpc from the centre) of the respective lens galaxy and are, thus, contaminated with its light. Since massive elliptical galaxies are empirically known to be characterised by a smoothly-varying light distribution and a very regular isophotal structure, a common practice to deal with this overlap is to fit the surface brightness of the lens galaxy with a parametric model and subtract it from the observed image. To this end, we apply and compare the performance of two techniques that have been successfully used in earlier studies of SLACS lenses -- the radial \textsc{b-spline} algorithm \citep{Bolton2006} and the empirical galaxy-fitting code \textsc{galfit} \citep{GALFIT_paper}.  

We perform the \textsc{b-spline} modelling using the implementation by \cite{Bolton2006}. This technique allows one to find the best-fitting coefficients $b_{mk}$ and $c_{mk}$ of the surface-brightness distribution $I(R,\phi)$ parametrized as follows:  
\begin{equation}
I(\rm{R},\theta) = \sum_{m,k} \big(b_{mk} \rm{cos}(m\phi) + c_{mk} \rm{sin} (m\phi)\big) f_{k}(\rm{R}),
\label{eq:b-spline}
\end{equation} 
where the radial dependence is modelled with a piecewise (linear, quadratic or cubic) polynomial function $f_{k}(R)$ defined on a chosen set of radial intervals $k$ and the angular dependence $\phi$ is fitted with a chosen number of multipole orders $m$. We set the radial interval breakpoints every 0.2 arcsec and fit $f_{k}(R)$ with piecewise-defined cubic polynomials. To model the angular dependence, we begin with the default configuration including the $m=0$ (monopole), $m=1$ (dipole) and $m=2$ (quadrupole) modes. If substantial angular structure is still found in the residual (data-model) image, we iteratively add further multiple orders, i.e. $m=4$ (octopole) and higher even terms, until the reduced $\chi^{2}$-statistic is minimized.

As an alternative, we apply the empirical galaxy-fitting technique \textsc{galfit} \citep{GALFIT_paper} and fit the lens light with a S\'ersic profile \citep{Sersic1963}, which is empirically known to provide a good fit to the surface-brightness distribution of observed massive elliptical galaxies. We iteratively add more S\'ersic components if justified by the residual pattern or a high value of the reduced $\chi^{2}$-statistic. In some cases, we improve the model by additionally fitting the diskiness/boxiness of the isophotes.

In both approaches, we exclude from the fit all pixels that, for various reasons, should not be taken into account during the galaxy-fitting procedure. More specifically, we mask out all pixels overlapping with the lensed images, nearby satellite galaxies, stellar streams and other astronomical objects not associated with but close in projection to the lens galaxy. In order to generate the mask, we first use the \textsc{ds9} software\footnote{http://ds9.si.edu/site/Home.html} to manually outline these features with a polygon. Subsequently, we determine the set of pixels contained inside this polygon by means of the \textsc{ds9poly} and the \textsc{fillpoly} software (freely available on the \textsc{galfit} webpage\footnote{http://users.obs.carnegiescience.edu/peng/work/galfit/galfit.html}). Finally, we make use of the \textsc{Pyraf}-task \textsc{badpiximage} to create a \textsc{fits}-image representing the mask. This initial mask is in some cases adjusted in the course of the galaxy-fitting procedure, in order to exclude additional faint features revealed in the residual image. 

\begin{figure*}
 \centering
 \includegraphics{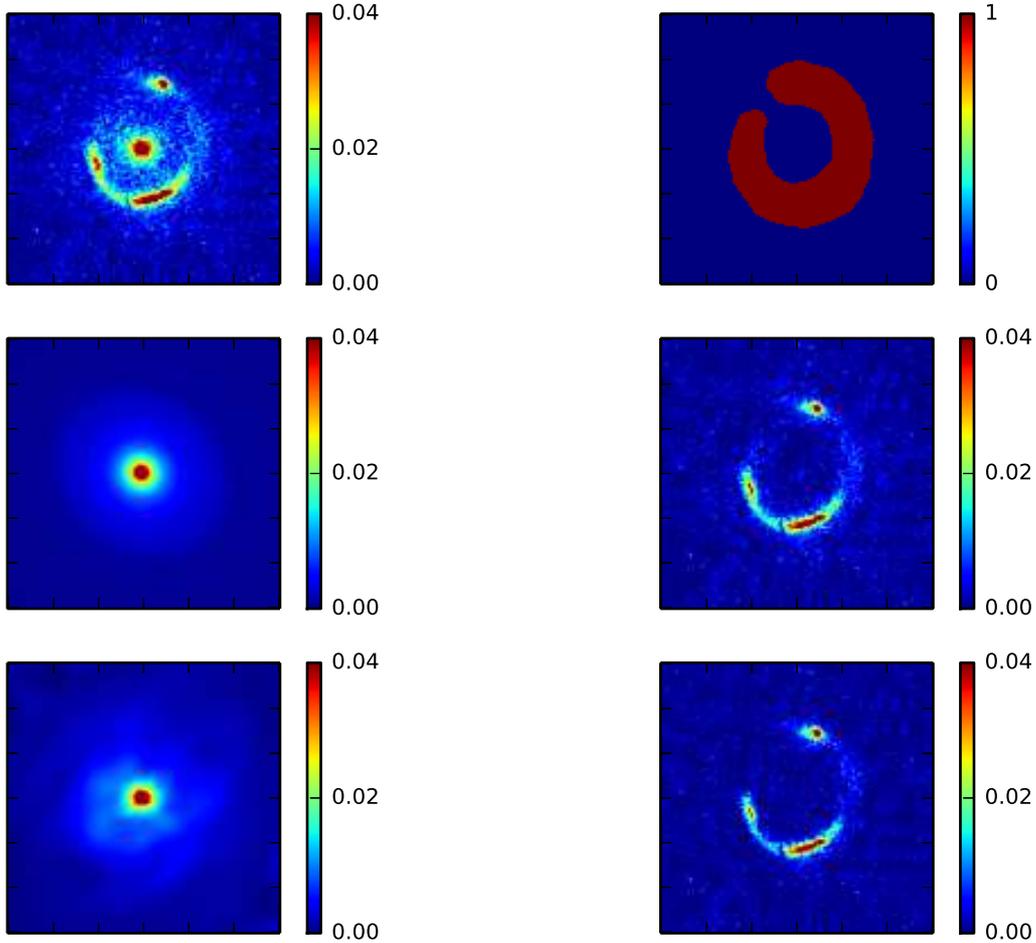}
 \caption{Lens-galaxy subtraction for SDSS J0737+3216 performed using alternatively \textsc{galfit} or the \textsc{b-spline} algorithm. \textit{Top row}: the observed HST-image in the F390W-band \textit{(left panel)} and the mask applied to exclude the lensed images from the fit \textit{(right panel)}. \textit{Middle row}: the best-fitting \textsc{galfit} model of the surface-brightness distribution in the lens galaxy \textit{(left panel)} and the lens-galaxy-subtracted residual image \textit{(right panel)}. \textit{Bottom row}: the best-fitting \textsc{b-spline} model \textit{(left panel)} and the corresponding residual image \textit{(right panel)}. The measured flux is expressed in units of electrons per second.}
\label{fig:Galfit_bspline}
\end{figure*}

In each case, the best-fitting model of the surface-brightness distribution in the lens galaxy is finally interpolated over the masked regions and subtracted from the original image. As an example, in Fig.~\ref{fig:Galfit_bspline}, we present the entire procedure of the lens-galaxy subtraction for the lens systems SDSS J0737+3216. The figure shows the original HST-image, the applied mask, the best-fitting \textsc{galfit} and \textsc{b-spline} models, and the respective galaxy-subtracted images. 

As a result, we find that for most of the lens systems in our sample \textsc{b-spline} provides a very good fit to the central region of the lens galaxy, however, the inferred global surface-brightness model has an irregular shape, deviating from our empirical expectations for massive elliptical galaxies. \textsc{galfit}, on the other hand, provides more realistic models, but it very often yields significant galaxy-core residuals, even when fitting multiple S\'ersic components. Thus, both techniques need to be applied with caution. The flexibility of \textsc{b-spline} allows to fit features that are not well described by standard parametric functions, but it might also lead to unrealistic models. To the contrary, fitting empirically-based parametric functions yields a reasonable solution in most cases, but might result in a poor fit if the galaxy deviates from the assumed typical morphology. In Section~\ref{Section:lens_subtraction_effect}, we investigate the effect of \textsc{galfit} and \textsc{b-spline} on the resulting power spectrum of the residual image and show that the two lens-galaxy-subtraction techniques lead to almost identical results in this respect. Since our objective is to estimate the surface-brightness contribution of the lens galaxy in the region overlapping with the lensed images and not necessarily to obtain the best estimate of the overall surface-brightness distribution, we choose \textsc{galfit} as the preferred method for the purpose of our analysis.

\begin{figure*}
\centering
 \includegraphics[scale=0.29]{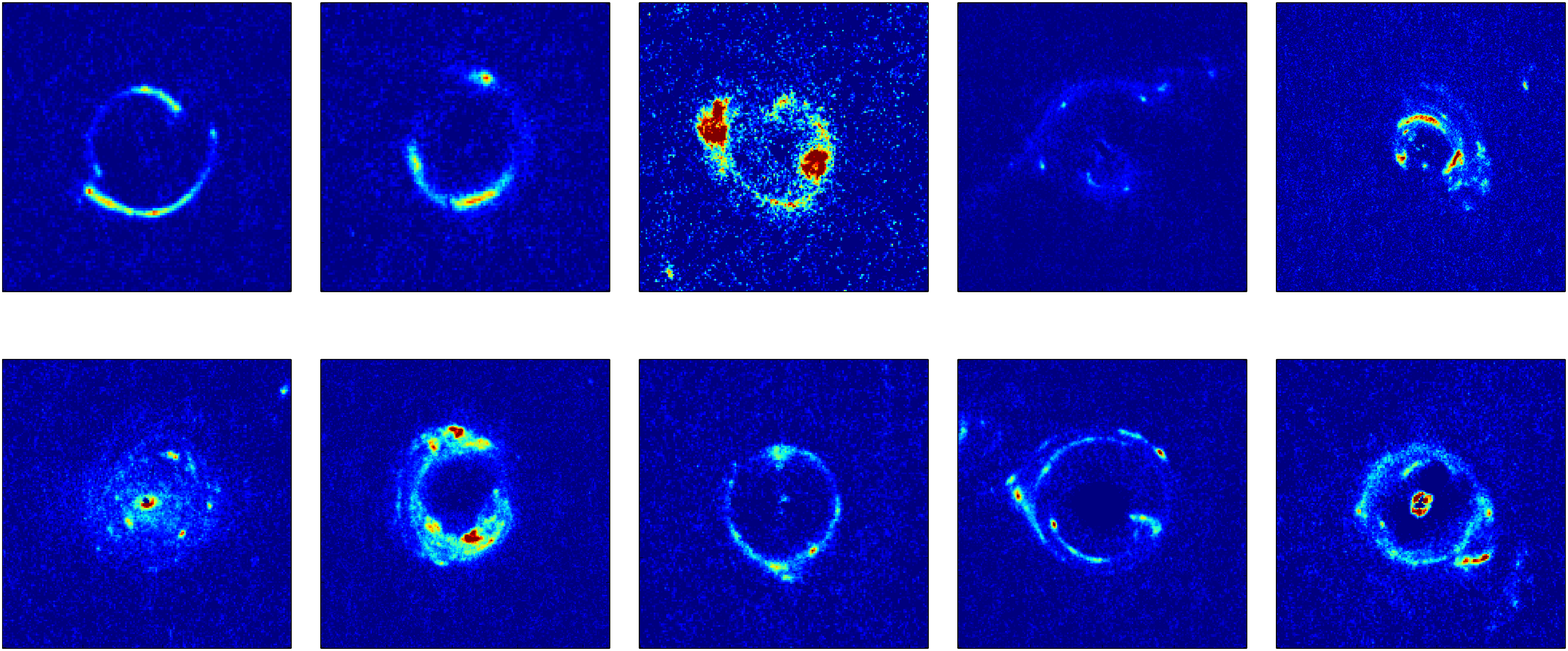}
 \caption{The lens-galaxy-subtracted images for all lens systems in our SLACS sub-sample, obtained using \textsc{galfit}. The significant galaxy-core residuals in some of the images might point towards the presence of a cusp or the central star formation in the (massive elliptical) lens galaxy. For the purpose of this study, we exclude these from further analysis and perform the lens modelling only for the remaining four lens systems having a relatively simple geometry and no substantial galaxy-core residuals.}
\label{fig:sample_galaxy_subtracted}
\end{figure*}

Fig.~\ref{fig:sample_galaxy_subtracted} shows the final galaxy-subtracted images, obtained using \textsc{galfit}. We note the presence of significant galaxy-core residuals in the majority of the images, which might point towards a cusp or central star formation in the (massive elliptical) lens galaxy \citep[see e.g.][]{Kaviraj_ellipticals_SF}. A viable way of mitigating this issue would be to either model the galaxy core separately and only then search for the best-fitting global model or to entirely exclude it from the fit. For the purpose of this study, however, we exclude these problematic lens systems from further analysis and perform the lens modelling only for the remaining four systems -- SDSS J0252+0039, SDSS J0737+3216, SDSS J1430+4105 and SDSS J1627--0053 -- with a relatively simple geometry and no substantial galaxy-core residuals.
 


\subsection{Smooth lens modelling} 
\label{Section:Lens_Model_UV_sample}

We apply the adaptive grid-based Bayesian smooth lens modelling technique by \cite{VegettiKoopmans2009} to model each selected lens system under the tentative assumption that the mass in the lens galaxy is distributed smoothly. More specifically, we assume its surface mass density to be well described by the power-law-elliptical-mass-distribution model \citep[PEMD,][]{Barkana1998}, with the convergence parametrized according to the convention used in \cite{VegettiKoopmans2009}:
\begin{equation}
\kappa(x,y) = \frac{b \ (2-\frac{\gamma}{2}) \ q^{\gamma-3/2}}{2(x^{2}q^{2}+y^{2})^{(\gamma-1)/2}}.
\label{eq:smooth_lens_convergence}
\end{equation} 
The model parameters are the lens strength $b$, the (minor to major) axis ratio $q$ and the (three-dimensional) mass-density slope $\gamma$ ($\gamma=2$ in the isothermal case). Moreover, the mass model is rotated and translated to fit the position angle of the major axis $\theta$ (measured with respect to the original telescope rotation) and the centroid location in the lens plane $x_{0}$ and $y_{0}$. In addition, we model the lensing effect of possible companion objects in the vicinity of the lens galaxy as an external shear field characterised by the shear strength $\Gamma$ and its position angle $\Gamma_{\theta}$. 

The lens-modelling code allows us to find the best-fitting parameter values of this PEMD-plus-external-shear macro model and, simultaneously, reconstruct the unlensed pixellated surface-brightness distribution of the source galaxy (on an adaptive grid in the source plane), which combined together most accurately reproduce the observed lensed images. However, there are several alternative ways to perform the mapping of pixels and flux values between the lens- and the source plane. Firstly, the resolution of the pixellated source reconstruction can be chosen by setting the value of a parameter referred to as $n$ which determines the linear size of a square in the lens plane out of which only the central pixel is cast back to the source plane. For example, if $n = 3$, only one pixel out of each contiguous $3\times3$-pixel area is used to create the reconstruction grid in the source plane, while $n = 1$ corresponds to casting back every single pixel. Note, however, that all pixels are used in the comparison between the data and the model, independently of the chosen $n$. Secondly, it is possible to apply different forms of source-grid regularisation, such as an adaptive or non-adaptive, variance, gradient or curvature regularisation \citep[see e.g.][]{Suyu_Regularization}. The optimal choice of the source-grid resolution and the form of regularisation depends on the level of structure in the source galaxy as well as on the signal-to-noise ratio of the data, and is usually made based on the highest value of the marginalized Bayesian evidence.

\begin{figure*}
 \includegraphics[scale=0.62]{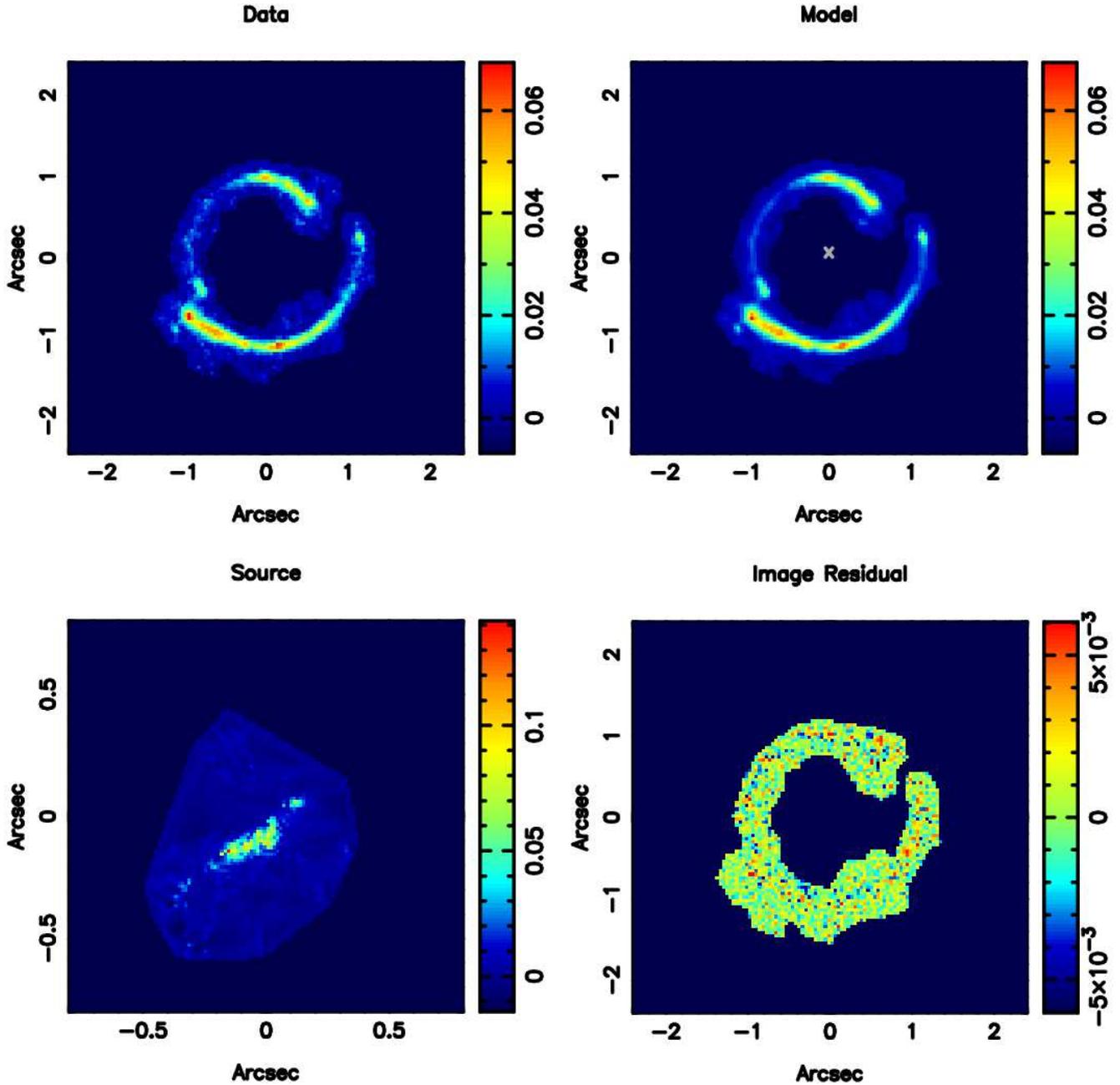}
 \centering
 \caption{Best-fitting power-law-elliptical-mass-distribution (PEMD) plus-external-shear macro model of SDSS J0252+0039 in the HST/WFC3/F390W-filter, inferred by means of the adaptive and grid-based Bayesian lens-modelling technique of \citet{VegettiKoopmans2009}. \textit{Top row}: the lens-galaxy subtracted image overlaid with a mask, used as input for the smooth lens modelling \textit{(left panel)} and the reconstructed smooth-lens model of the lensed images \textit{(right panel)}. \textit{Bottom row}: the unlensed surface-brightness distribution of the source galaxy reconstructed on an adaptive grid in the source plane (more specifically a Delaunay triangulation regridded to pixels for plotting purposes) with the source-grid resolution $n=1$, i.e. casting back every pixel from the image plane to the source plane \textit{(left panel)}, and the residual (data - model) image showing the remaining surface-brightness fluctuations that are not explained by the smooth-lens model \textit{(right panel)}. The reconstructed parameter values of the best-fitting smooth lensing potential as well as the chosen modelling options can be found in Table~\ref{tab:smooth_lens_parameters_UV_sample}.}
 \label{fig:J0252_smooth_lens}
\end{figure*}

\begin{figure*}
 \includegraphics[scale=0.69]{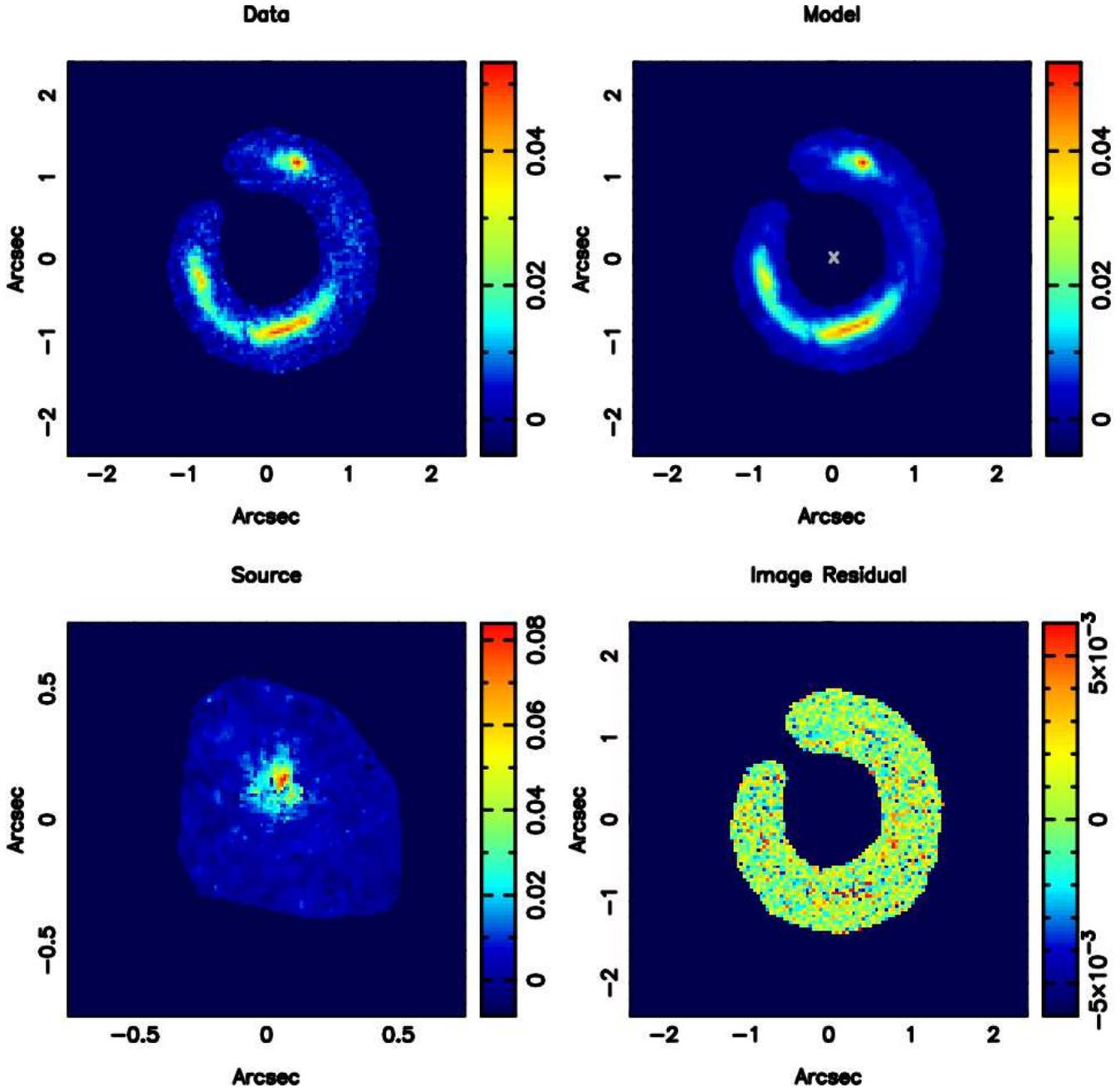}
 \centering
 \caption{Idem as in Fig. \ref{fig:J0252_smooth_lens} for the lens system SDSS J0737+3216.}
 \label{fig:J0737_smooth_lens}
\end{figure*}

\begin{figure*}
 \includegraphics[scale=0.68]{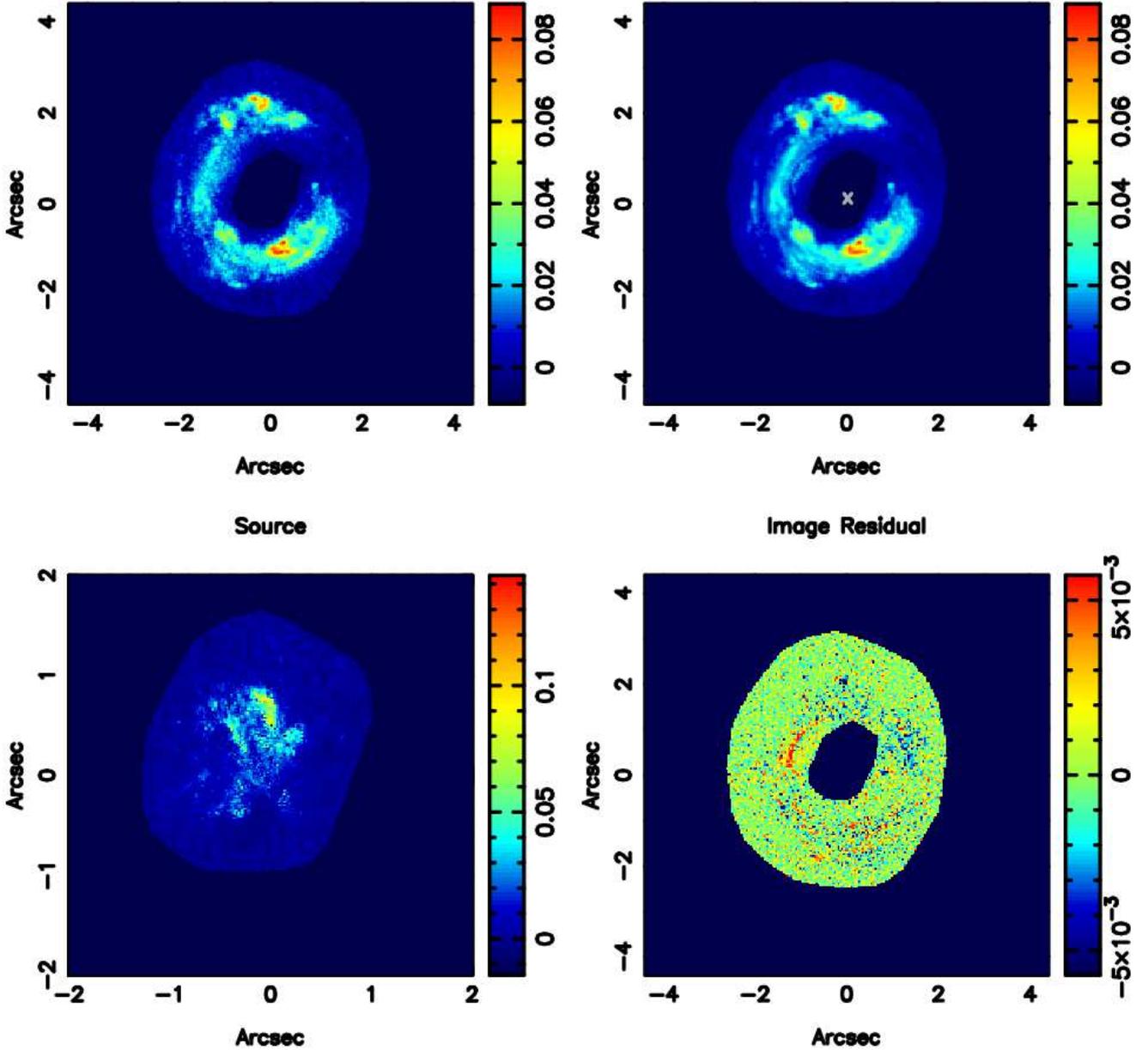}
 \centering
 \caption{Idem as in Fig. \ref{fig:J0252_smooth_lens} for the lens system SDSS J1430+4105.}
 \label{fig:J1430_smooth_lens}
\end{figure*}

\begin{figure*}
 \includegraphics[scale=0.62]{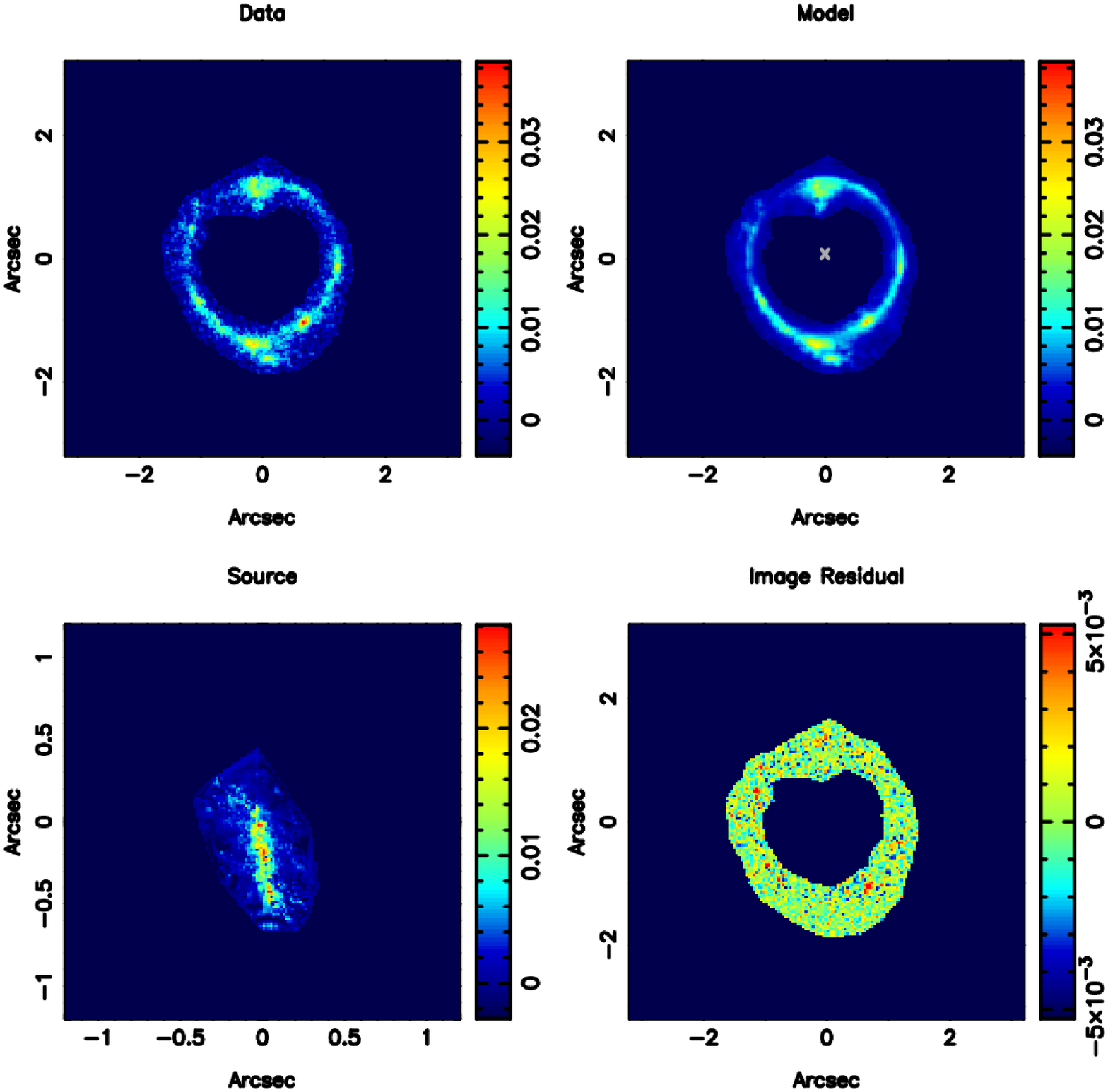}
 \centering
\caption{Idem as in Fig. \ref{fig:J0252_smooth_lens} for the lens system SDSS J1627--0053.}
\label{fig:J1627_smooth_lens}
\end{figure*}

We obtain smooth-lens models with the highest Bayesian evidence when choosing the highest resolution ($n=1$) and an adaptive gradient regularisation for all four analysed lens systems (but see Section~\ref{Section:smooth_lens_PS} for a discussion on the overfitting problem). Table~\ref{tab:smooth_lens_parameters_UV_sample} presents the inferred parameter values of the best-fitting PEMD-plus-external-shear macro models, in comparison to the earlier F814W and F555W reconstructions carried out by \citet{Vegetti2014} (where available). Figs.~\ref{fig:J0252_smooth_lens} to \ref{fig:J1627_smooth_lens} depict the respective modelled data, the inferred best-fitting model and the reconstructed surface-brightness distribution of the source galaxy on an adaptive grid in the source plane. Moreover, each of these panels shows the resulting residual image representing the deviation of the observed lensed images from the best-fitting model. The key idea of our approach is that, apart from noise, these residual surface-brightness fluctuations might be caused by perturbations in the lensing potential due to small-scale mass structure in the lens galaxy.

\begin{table*}
\begin{center}
\caption{Parameter values of the best-fitting power-law-elliptical-mass-distribution (PEMD) plus-external-shear macro model for four lens galaxies from our SLACS sub-sample, selected for further analysis due to a relatively simple geometry and no substantial galaxy-core residuals, in comparison to earlier F814W and F555W reconstructions performed by \citet{Vegetti2014}. The models are based on lens-galaxy-subtracted HST/WFC3/F390W-images obtained using alternatively \textsc{galfit} or \textsc{b-spline}. The mass distribution of the lens galaxy is modelled with the following set of free parameters: the lens strength $b$, the position angle $\theta$ (with respect to the original telescope rotation), the axis ratio $q$, the (three-dimensional) mass-density slope $\gamma$, the external shear strength $\Gamma$ and its position angle $\Gamma_{\theta}$. Additionally, we quote the source-grid resolution $n$ and the type of source-plane regularisation (Reg.) that lead to models with the highest marginalized Bayesian evidence. \rm{G} indicates the gradient regularisation, \rm{C} the curvature regularisation and \textit{adp} an adaptive regularisation. The typical statistical errors of the parameter values are of the order $10^{-1}$ for the angles and $10^{-3}$ for the remaining parameters.}
\label{tab:smooth_lens_parameters_UV_sample}
\begin{tabular}{ccccccccccc} 
		\hline
		Lens system & Filter & Lens-galaxy subtraction & $b [\rm{arcsec}]$ & $\theta$ [deg.] & $q$ & $\gamma$ & $\Gamma$ & $\Gamma_{\theta}$ [deg.]& $n$ & Reg.\\
		\hline
		J0252+0039 & F390W & Galfit& 0.996 & 150.1 & 0.978 & 2.066 & -0.015 & 81.4 & 1& $G_{adp}$\\
        & F390W & b-spline& 0.996 & 149.2 & 0.978 & 2.066 & -0.015 & 81.4 & 1 & $G_{adp}$\\
		 & F814W & b-spline & 1.022 & 26.2 & 0.943 & 2.047 & 0.009 & 101.8 & 1 & $G_{adp}$\\
         
         J0737+3216 & F390W & Galfit& 0.926 & 66.1 & 0.862 & 2.110 &0.066  & 72.8 &1 &$G_{adp}$\\
         & F390W & b-spline& 0.933 & 66.9 & 0.869 & 2.102 & 0.062 & 72.2 & 1 & $G_{adp}$\\
		 & F814W & b-spline& 0.951 & 78.3 & 0.705 & 2.066 & 0.050 & 100.8 & 1 & $C$\\
         & F555W & b-spline& 0.951 & 77.2 & 0.709 & 2.073 & 0.052 & 102.5 & 2 & $G_{adp}$\\
         
         J1430+4105 & F390W & b-spline& 1.527 & 85.4 & 0.647 & 2.073 &0.029  & 139.2 &1 &$G_{adp}$\\
         & F814W & b-spline& 1.484 & 61.5 & 0.710 & 2.048 & 0.051 & 128.6 & 1 & $C$\\
         
         J1627--0053 & F390W & b-spline& 1.217 & 16.9 & 0.856 & 2.006 &0.008  & 101.3 &1 &$G_{adp}$\\
         & F390W & Galfit& 1.218 & 17.0 & 0.855 & 2.007 &0.008  & 101.5 &1 &$G_{adp}$\\
         & F814W & b-spline& 1.229 & 14.3 & 0.912 & 1.998 & 0.004 & 80.0 & 2 & $C_{adp}$\\
         & F555W & b-spline& 1.212 & 14.2 & 0.869 & 2.058 & 0.014 & 87.8 & 2 & $C_{adp}$\\
		\hline
\end{tabular}
\end{center}
\end{table*}

\subsection{Power-spectrum analysis of the residual surface-brightness fluctuations}
\label{Section:PS}

The residual images of all investigated lens systems reveal surface-brightness fluctuations that cannot be explained by the assumed smooth PEMD-plus-external-shear macro model of the mass distribution in the lens galaxy, as demonstrated in Figs.~\ref{fig:J0252_smooth_lens} to \ref{fig:J1627_smooth_lens}. In this section, we estimate the variance of these residuals as a function of their spatial scale, i.e. the power spectrum, following the approach proposed by \cite{Sander_thesis} and \cite{Saikat}. Moreover, we investigate how the different choices made in the process of the lens-galaxy subtraction and the smooth lens modelling affect the measured residual power spectrum.

\subsubsection{Power-spectrum measurement}
\label{Section:PS_measurement}

The goal of this power-spectrum analysis is to decompose the residual surface-brightness fluctuations ${\delta I}(x)$ into modes with different length scales $\lambda$, expressed in terms of the corresponding wavenumbers $k$. We note that in this study we follow the convention in which the wavenumber is equal to the reciprocal length scale: 
\begin{equation} k\equiv\lambda^{-1} \label{eq:k_definition} \end{equation} and is measured in $\mathrm{arcsec^{-1}}$. 

We calculate the residual power spectrum for each modelled lens system individually, within the respective mask covering the lensed images. For this, we set the flux values of all pixels located outside the mask to zero and compute the two-dimensional discrete Fourier transform (DFT) of the masked residual image using the Python package \textsc{numpy.fft} \footnote{https://docs.scipy.org/doc/numpy/reference/routines.fft.html}. The squared magnitudes of the obtained (complex-valued) Fourier coefficients, assigned to the individual pixels of the Fourier-transformed residual image, yield the two-dimensional power spectrum of the residual surface-brightness fluctuations. We further assume the residuals to be isotropic and average this two-dimensional power spectrum along a set of ten equidistant concentric annuli covering the full Fourier-transformed image. The resulting one-dimensional azimuthally-averaged power spectrum $P(k)$ constitutes the final statistic allowing us to perform a statistical comparison of the residual surface-brightness fluctuations resulting from different models. 

\subsubsection{Effect of the lens-galaxy subtraction}
\label{Section:lens_subtraction_effect}

\begin{figure}
\includegraphics[width=\columnwidth]{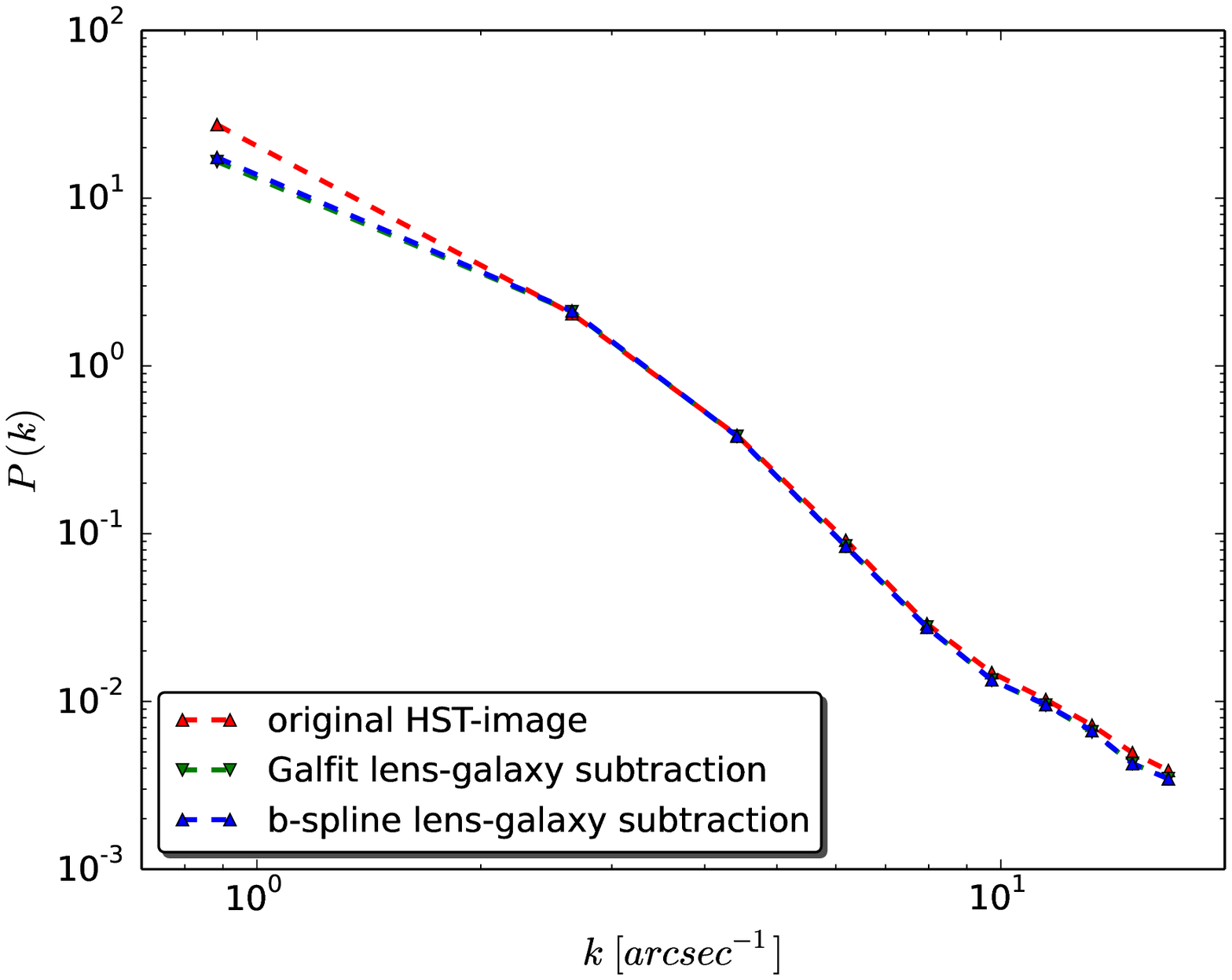}
\includegraphics[width=\columnwidth]{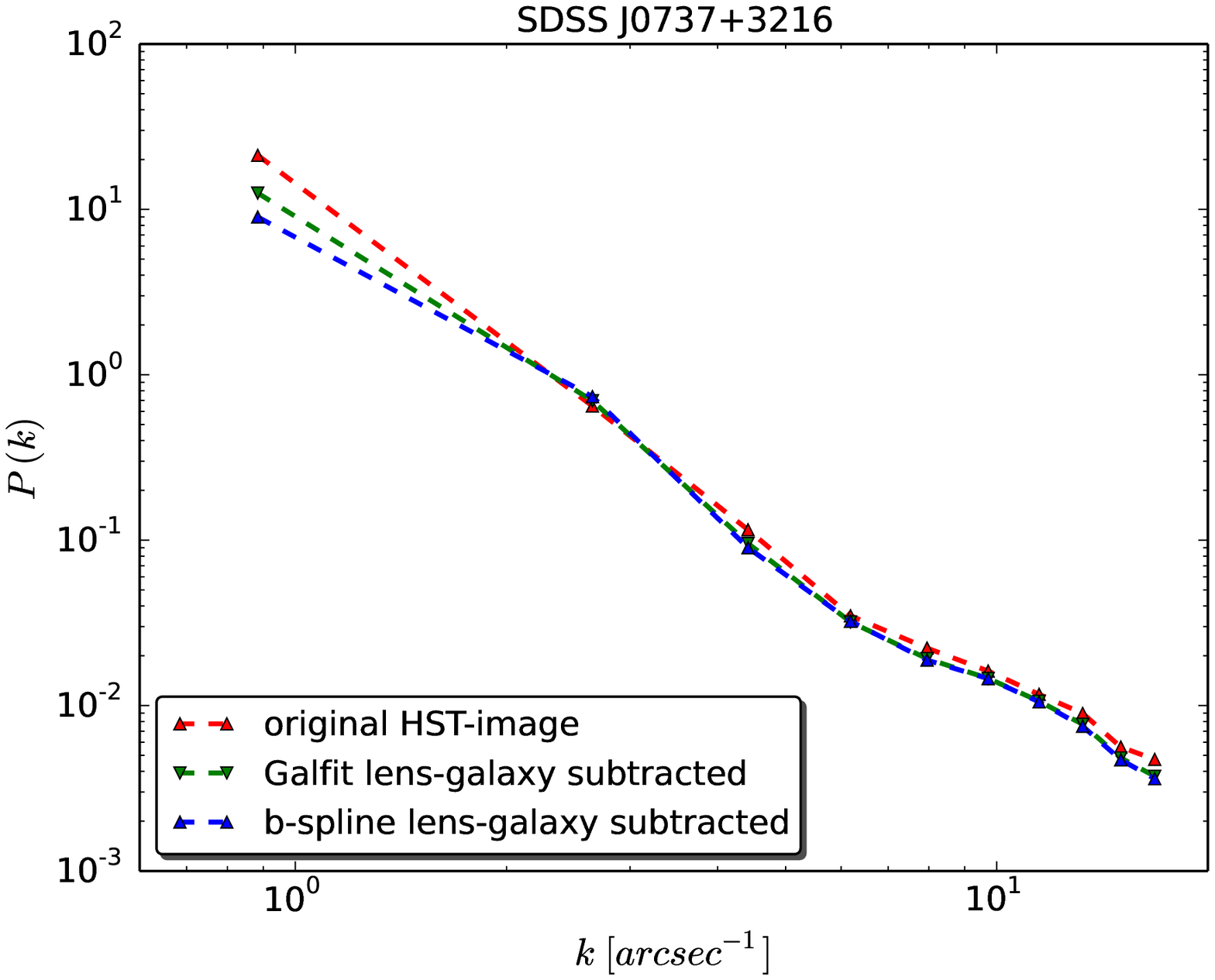}
\includegraphics[width=\columnwidth]{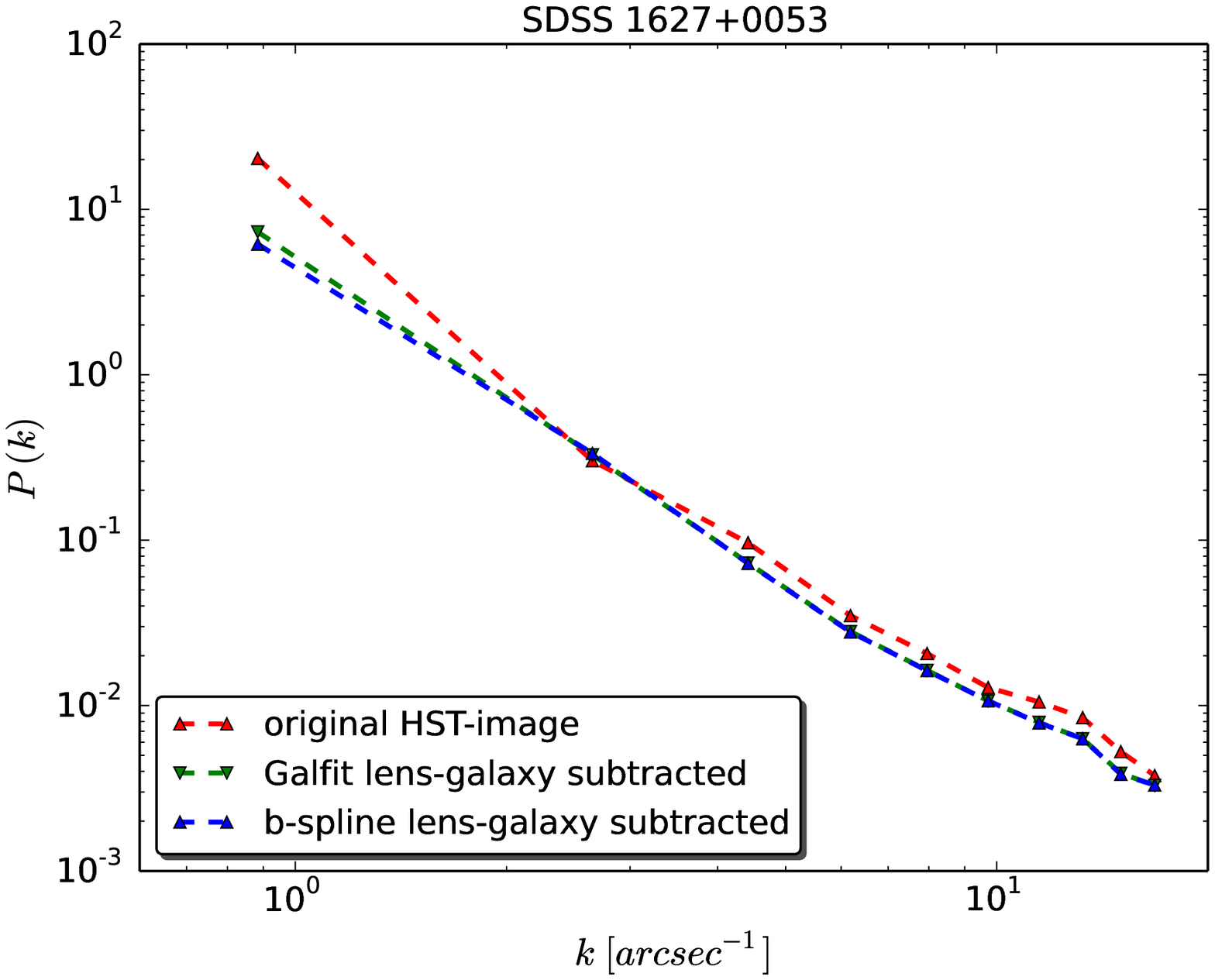}
\caption{Effect of the lens-galaxy subtraction, performed using alternatively \textsc{galfit} or \textsc{b-spline}, on the power spectrum of surface-brightness fluctuations in the galaxy-subtracted images for SDSS J0252+0039, SDSS J0737+3216 and SDSS 1627+0053.}
\label{fig:PS_galaxy_subtraction}
\end{figure}

In order to assess whether the choice of the lens-galaxy-subtraction technique, i.e. \textsc{galfit} or \textsc{b-spline}, might be a source of a systematic bias, we calculate the azimuthally-averaged power spectrum of the residual images after the lens-galaxy subtraction using both methods for SDSS J0252+0039, SDSS J0737+3216 and SDSS 1627+0053. We note that we were not able to obtain a good \textsc{galfit} model for SDSS J1430+4105 and, thus, the system is omitted from this comparative power-spectrum analysis. The outcome of this test, presented in Fig.~\ref{fig:PS_galaxy_subtraction}, shows that these two galaxy-subtraction techniques lead to almost identical power spectra. As expected, by removing a smooth large-scale surface-brightness component, the lens-galaxy subtraction reduces the measured power spectrum only on the largest spatial scales (i.e. smallest $k$-values). 

We conclude that if the power-spectrum analysis is performed in the ultra-violet band, where the surface-brightness distribution of elliptical galaxies peaks strongly in the centre and decreases very quickly towards the outskirts, the choice of the galaxy-subtraction technique does not appreciably affect the final results. In a different band, however, especially in the infrared where the lens galaxy dominates the surface brightness in the region overlapping with the lensed images, the choice of the lens-light model might significantly alter the analysis outcome. We plan to investigate this issue in our future paper.

\subsubsection{Effect of the smooth lens modelling}
\label{Section:smooth_lens_PS}

As stated in Section~\ref{Section:Lens_Model_UV_sample}, we obtain smooth-lens models with the highest Bayesian evidence when the lens modelling is performed with the highest resolution ($n=1$) and an adaptive gradient source-grid regularisation. However, it turns out that this choice leads to a power spectrum of the residual surface-brightness fluctuations lying at or below the noise level for all four modelled lens systems (see Section~\ref{Section:Noise_correction} for the estimation of the noise power spectrum). This means that all surface-brightness anomalies due to the hypothetical small-scale mass structures in the lens galaxy and even a substantial fraction of the background noise have been modelled as spurious structure in the intrinsic surface-brightness distribution of the source galaxy. This problem is generally known as overfitting (i.e. modelling of the inherent noise present in the data) and arises when the number of free parameters in a model is much larger than the number of the imposed constraints. In our study, it might also arise from an incorrect or incomplete macro model.

\begin{figure}
\centering
\includegraphics[width=\columnwidth]{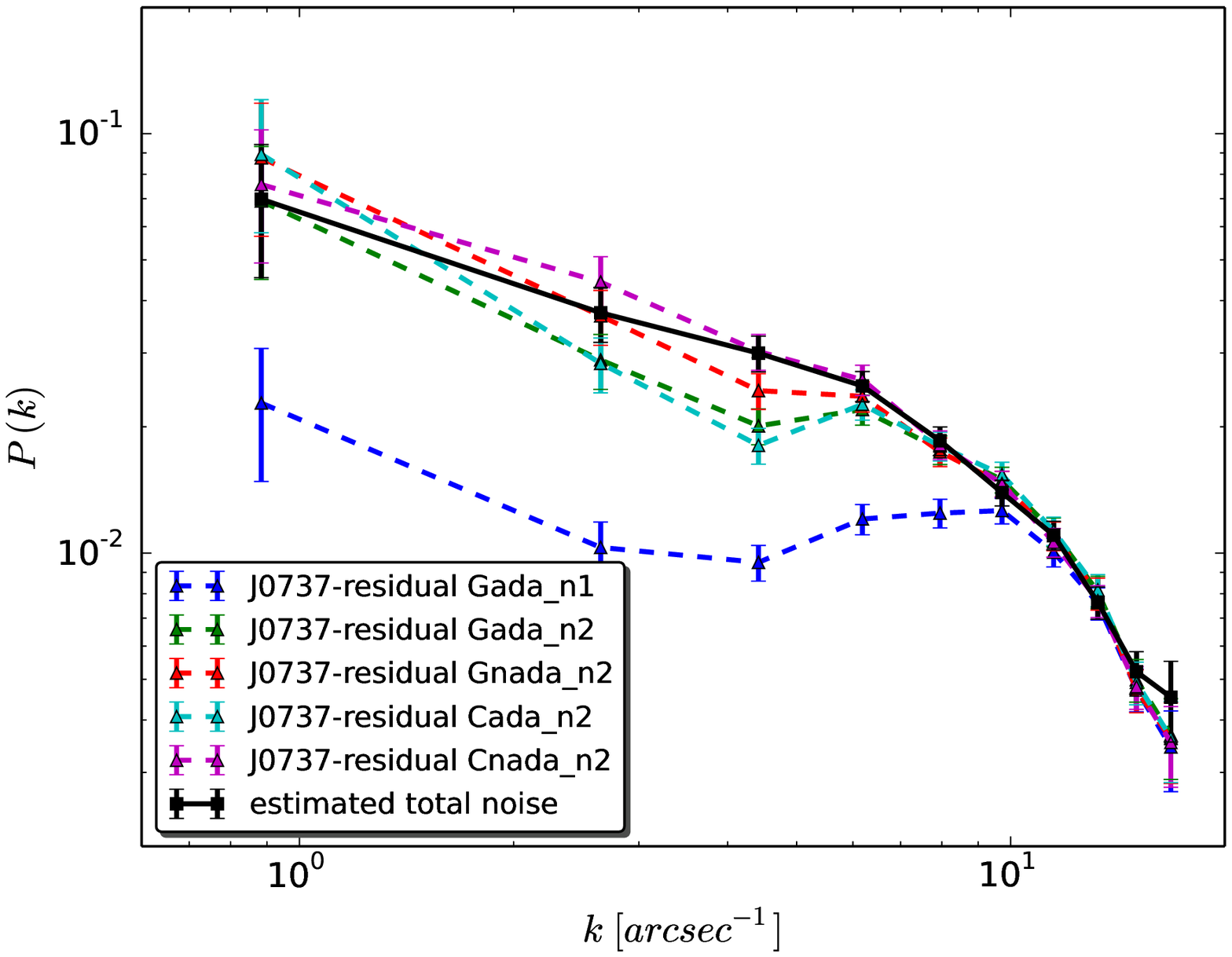}
\caption{Power spectrum of the residual surface-brightness fluctuations remaining in the lensed images of SDSS J0737+3216 after the lens-galaxy subtraction and the smooth lens modelling for varying levels of the source-grid resolution and different types of the source-grid regularisation. In the legend, \textit{G} indicates the gradient regularisation, \textit{C} the curvature regularisation, \textit{ada} an adaptive and \textit{nada} a non-adaptive source-plane regularisation. The symbols \textit{n1} and \textit{n2} stand for the source-grid resolution corresponding to $n=1$ and $n=2$, respectively. The computed power spectrum lies below the noise level for all models with the highest resolution ($n=1$ and $n=2$), irrespective of the chosen form of regularisation.}
\label{fig:J0737_smooth_models_n2}
\end{figure}

In order to investigate this issue, we perform tests with a lower resolution and different forms of the source-grid regularisation. As an example, Fig.~\ref{fig:J0737_smooth_models_n2} shows the effect of these different options on the power spectrum of the residual surface-brightness fluctuations measured in the lensed images of the lens system SDSS J0737+3216. As is apparent from this figure, the computed residual power spectrum lies at or below the noise level for all models with $n=1$ or $n=2$, irrespective of the chosen form of regularisation. More generally, we find that for almost all investigated lens systems even the choice of $n=2$ still leads to the residuals lying below the noise level. From this, we conclude that if the smooth lens modelling in the U-band is carried out with the highest resolution (i.e. $n=1$ or $n=2$) and a relatively tight mask, as in earlier F814W and F555W reconstructions (see Table~\ref{tab:smooth_lens_parameters_UV_sample}), the inversion problem to be solved is underconstrained and degenerate.

\begin{figure} 
 \includegraphics[width=\columnwidth]{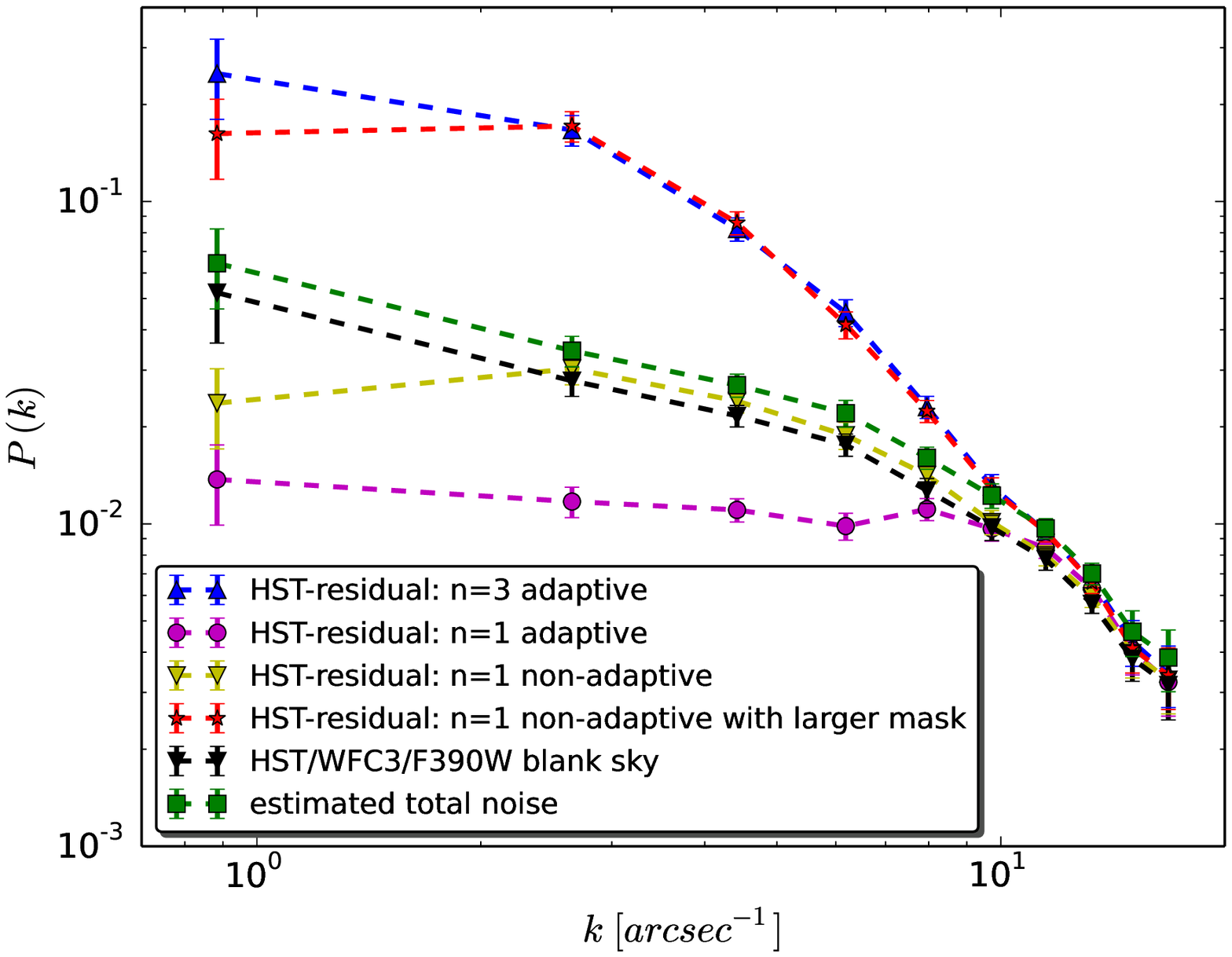}
 \caption{Power spectrum of the residual surface-brightness fluctuations remaining in the lensed images of SDSS J0252+0039 after the smooth lens modelling. The lens modelling is performed alternatively with the highest source-grid resolution ($n = 1$), using either an adaptive (\textit{magenta line}) or a non-adaptive (\textit{yellow line}) source-grid regularisation, or with a lower source-grid resolution ($n = 3$; \textit{blue line}) chosen for this study to prevent overfitting. As an alternative solution, the \textit{red line} shows the effect of using a larger mask in the smooth lens modelling combined with $n = 1$ and a non-adaptive source-grid regularisation (for consistency reasons the power spectrum is calculated within the original tight mask). The sky-background power spectrum \textit{(black line)} is estimated based on a sample of twenty blank-sky regions (overlaid with the original tight mask) in the proximity to the lens. The estimated total noise power spectrum \textit{(green line)} accounts for the additional flux-dependent (Poisson) shot noise in the science image.}
 \label{fig:HSTn1vsNoise}
\end{figure}

On the other hand, higher $n$-values might deteriorate the sampling of the lensed images and, consequently, diminish the accuracy of the source reconstruction. Hence, a balance needs to be found between the possible over- and under-fitting. In the present study, we mitigate the overfitting problem by lowering the resolution of the source reconstruction even further to $n=3$ (i.e. only one pixel out of each contiguous $3\times3$-pixel area is used to create the reconstruction grid in the source plane) while keeping fixed the best-fitting parameter values of the smooth lensing potential inferred with the highest resolution ($n=1$). As is shown in Fig.~\ref{fig:HSTn1vsNoise} for the lens system SDSS J0252+0039, this approach allows us to prevent overfitting and leads to the residual power spectrum lying at or above the noise level for all $k$-bins. In Section~\ref{Section:Performance_test}, we additionally demonstrate that the choice of $n=3$ enables us to successfully recover the known true surface-brightness anomalies in a mock lens system mimicking SDSS J0252+0039.

Alternatively, the application of a larger mask including more noise-dominated pixels in the smooth-lens-modelling procedure (in combination with $n = 1$ and a high non-adaptive source-grid regularisation) might offer another solution to alleviate the overfitting problem in the source reconstruction. As apparent from Fig.~\ref{fig:HSTn1vsNoise}, both options lead to almost identical power spectra of the residual surface-brightness fluctuations in the lensed images (consistently calculated within the original tight mask), except for a small difference in the lowest analysed $k$-bin. Our preliminary tests confirm that this does not affect our final results (i.e. exclusion probabilities of the matter-power-spectrum models, see Paper II) significantly. However, since the respective lens models are obtained using different masks and, thus, cannot be considered as inferred from the same data set, a proper comparison in terms of the Bayesian evidence is not possible. We are planning to investigate this alternative in more detail in Paper III of this series (Bayer et al., in prep).

\subsection{Noise power spectrum analysis}
\label{Section:Noise_correction}

Besides possible surface-brightness anomalies due to mass structure in the lens galaxy, the residual surface-brightness fluctuations remaining in the lensed images after subtraction of the best-fitting smooth lens model are partially caused by the observational noise. In this section, we estimate the noise contribution to the measured power spectrum of the residual surface-brightness fluctuations. 

\subsubsection{The noise-sigma maps}
\label{Section:noise_sigma}

\begin{figure*}
\centering
\includegraphics[scale=1.0]{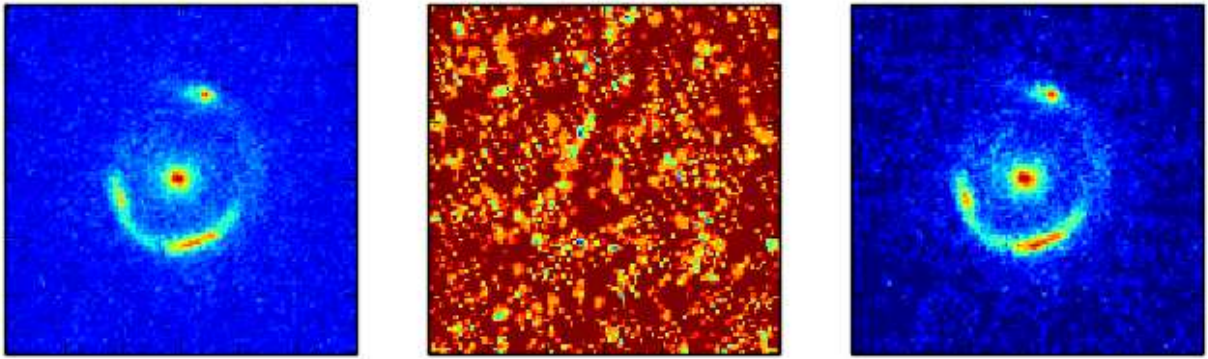}
\caption{The process of generating the noise-sigma map for the observed HST/WFC3/F390W-image of the lens system SDSS J0737+3216: the drizzled science image with photon counts $N$ \textit{(left panel)}, the inverse-variance weight map $W$ from the drizzling procedure \textit{(middle panel)} and the resulting noise-sigma map $\sigma_{n}$ (see equation \ref{eq:noise_sigma_map}, \textit{right panel}). The standard deviation of the sky background $\sigma_{\rm sky}$ is approximated by the value measured in a sample of close-by blank-sky cutouts with the same size as the science image.}
\label{fig:cutout_weight_noise_J0737}
\end{figure*}

Formally, the observational noise in our HST-imaging can be thought of as a random field. Each pixel is assigned a random variable representing the flux noise, with the expectation value equal to zero (after sky subtraction) and a flux-dependent variance. Two main contributions to this variance are the random fluctuations of the sky background and the Poisson-distributed photon-shot noise from the observed lens system. These are independent random processes, thus the total variance of the observational noise $\sigma_{n}^2$ in a given pixel can be expressed as the sum of the two variance components:
\begin{equation}
\sigma_{n}^2 = \sigma_{\rm sky}^2 + \sigma_{P}^2.
\label{eq:noise_sigma_map}
\end{equation} 
We approximate the standard deviation of the sky-background fluctuations $\sigma_{\rm sky}$ in the analysed images by the standard deviation of the flux values measured in a sample of close-by blank-sky cutouts $\sigma_{\rm sky}= 0.002 \ \mathrm{e^- \ sec^{-1}}$. The variance of the photon-shot noise $\sigma_{P}^2$, on the other hand, is a priori not known but, following the Poisson distribution, equal to the expected flux. The latter can be (due to the large number of counts in our imaging) approximated by the number of electrons per second $N$ measured in a given pixel (after the sky-background subtraction), weighted by the inverse-variance weight $W$ from the weight map computed in the process of drizzling:
\begin{equation}
\sigma_{P}^2 = N/W. 
\end{equation}
The estimated standard deviation of the total observational noise $\sigma_{n}$ in all individual pixels of our HST-images is finally presented in the form of noise-sigma maps. As an example, Fig.~\ref{fig:cutout_weight_noise_J0737} illustrates this procedure for the lens system SDSS J0737+3216. 

A noise-sigma map provides a complete description of the noise properties in an image, provided that the random flux fluctuations in the different pixels are statistically independent from each other. However, as we discuss in the next Section~\ref{Section:Noise_correlation_UV}, drizzled images are known to show noise correlations between adjacent pixels, which requires a more thorough noise analysis.


\subsubsection{Noise correlations due to drizzling}
\label{Section:Noise_correlation_UV}

Despite the fact that the individual pixels in raw HST/WFC3/F390W-images can (ideally) be considered independent, drizzled images are known to show noise correlations. In the process of drizzling, pixels from multiple dithered exposures are aligned and mapped (or informally \textit{drizzled}) onto a common output grid, based on the relative shift and rotation (i.e. \textit{dither}) of the respective exposure. The flux of each input pixel is then redistributed over all overlapping output pixels (according to the fractional overlap), which introduces correlations between adjacent pixels in the final drizzled image \cite[for a detailed discussion on noise correlations in drizzled images see][]{Casertano}.

\begin{figure*}
\centering
\includegraphics[width=\columnwidth]{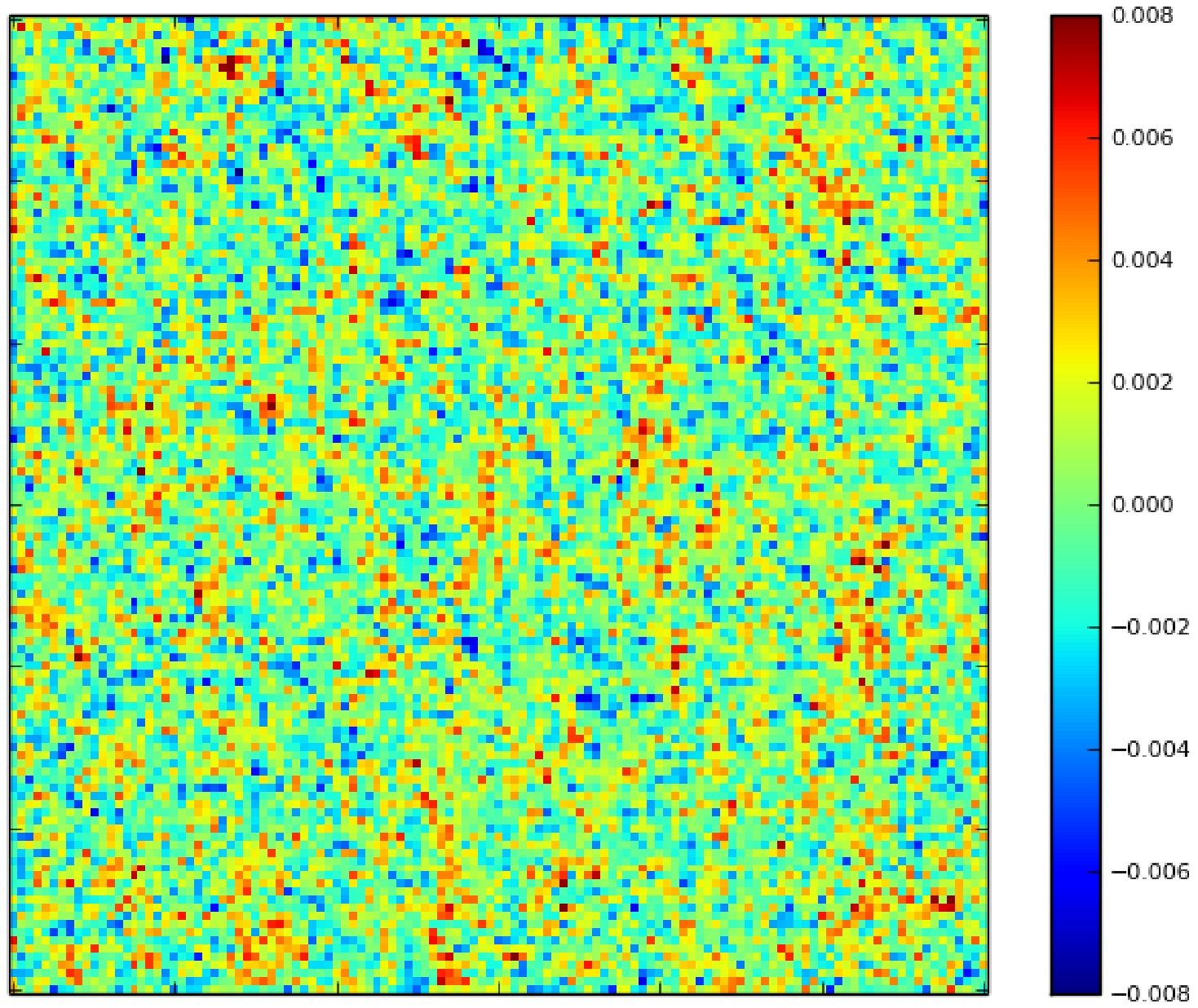}
\includegraphics[width=\columnwidth]{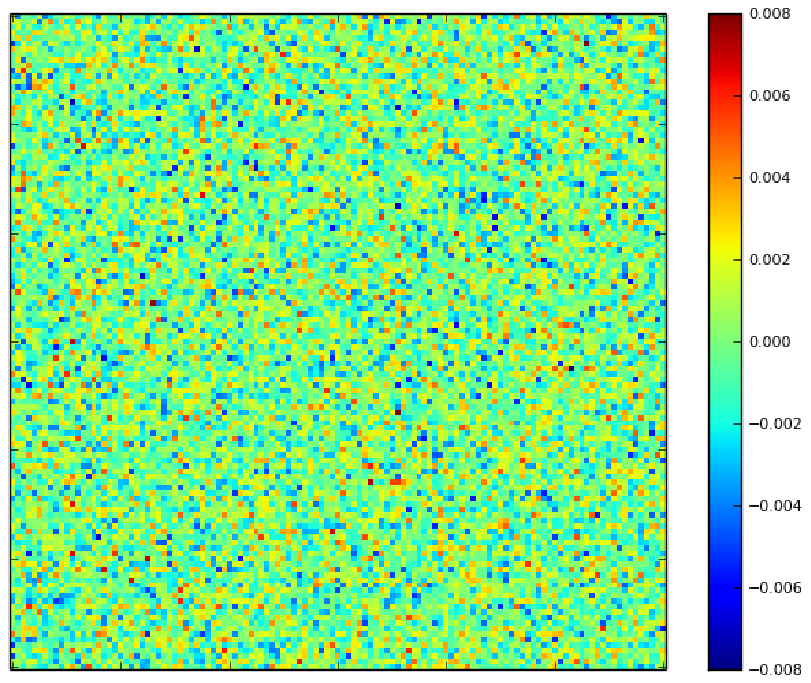}
\caption{Noise correlations introduced by drizzling (default configuration): a drizzled blank-sky cutout observed with HST/WFC3/F390W \textit{(left panel)} vs. a realisation of an uncorrelated Gaussian noise representing the true statistically independent fluctuations of the sky background \textit{(right panel)}. Both images have the same total variance of flux values, however, the drizzled image shows a distinct patchy correlation pattern.}
\label{fig:white_noise_ES_FITS}
\end{figure*}

In order to investigate the noise-correlation pattern in our data, we create a sample of drizzled blank-sky cutouts located in the proximity to each analysed lens system and quantify their statistical properties in terms of the azimuthally-averaged power spectrum. For consistency reasons, we match the size of these blank-sky fields to the size of the respective science image. As an example, Fig.~\ref{fig:white_noise_ES_FITS} depicts one of these drizzled blank-sky fields, located in proximity to the lens system SDSS J0252+0039, in comparison to a realisation of uncorrelated Gaussian noise with the same variance of the flux values. Whereas the latter represents the true statistically independent fluctuations of the sky background, the drizzled blank-sky image exhibits a distinct blotchy correlation pattern. 

\begin{figure}
\centering
 \includegraphics[width=\columnwidth]{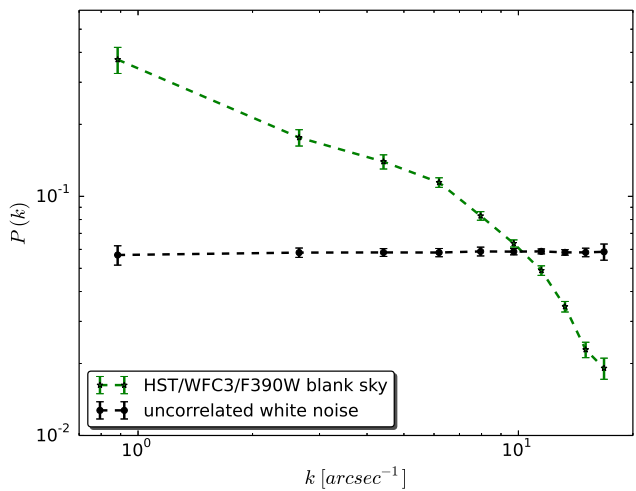}
 \caption{Azimuthally-averaged power spectrum of the sky background in drizzled HST/WFC3/F390W-images: the mean azimuthally-averaged power spectrum measured in a sample of twenty drizzled (default configuration) blank-sky regions located in proximity to SDSS J0252+0039 \textit{(green line)} vs. the flat power spectrum measured in a sample of twenty mock realisations of uncorrelated Gaussian noise (see Fig.~\ref{fig:white_noise_ES_FITS} for an example) with the same variance as in the observed blank-sky cutouts \textit{(black line)}.}
\label{fig:correlated_empty_sky_UV}
\end{figure}

Fig.~\ref{fig:correlated_empty_sky_UV} presents the mean power spectrum measured in a sample of twenty such blank-sky fields located in proximity to SDSS J0252+0039, in comparison to a sample of twenty realisations of uncorrelated Gaussian noise with the same total variance. Whereas the power spectrum of the uncorrelated Gaussian-noise realisations is flat (as expected), the power spectrum of the drizzled blank-sky cutouts is scale-dependent and carries a signature of the correlation pattern imposed by the drizzling procedure. The variance in the drizzled images is larger than in the Gaussian-noise realisations on large spatial scales, but smaller on small spatial scales (below 0.1 arcsec or 2.5 pixels).

However, a crucial feature of the \textit{Drizzle} algorithm is the possibility to improve the spatial resolution and reduce noise correlations in the final drizzled image by simultaneously decreasing the pixel scale of the output grid and shrinking the input pixels before mapping them onto the finer output grid \citep{Drizzle}. The pixel size of the output grid is controlled by the \textit{final pixscale} parameter, whereas the size of the shrunken input pixels, called \textit{drops}, is varied by means of the \textit{final pixfrac} parameter. The latter sets the ratio between the linear size of the drop and the original input pixel. The flux of each drop is then redistributed among overlapping output pixels with a weight proportional to the overlap. In comparison to the default configuration, in which both the output pixel and the drop have the same size as the original input pixels, i.e. \textit{final pixscale} = 0.04\,arcsec (for HST/WFC3/UVIS-imaging) and \textit{final pixfrac} = 1, the flux of the shrunken pixels is redistributed among fewer output pixels, which could help reduce the noise correlations between adjacent pixels. 

\begin{figure}
\centering
 \includegraphics[width=\columnwidth]{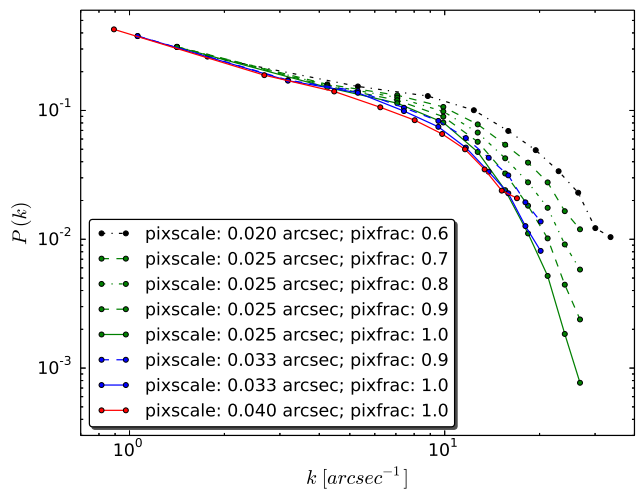}
 \caption{Effect of the drizzling parameters \textit{final pixscale} (linear size of the output pixels) and \textit{final pixfrac} (size of the \textit{drops}) on the azimuthally-averaged power spectrum of surface-brightness fluctuations in a selected blank-sky region located in proximity to SDSS J0252+0039 (see left panel of Fig.~\ref{fig:white_noise_ES_FITS}). The default configuration corresponds to \textit{final pixscale} = 0.04 arcsec and \textit{final pixfrac} = 1 \textit{(red line)}.}
 \label{fig:pixscale_pixfrac_UV}
\end{figure}

To test this possibility of reducing the noise correlations in our HST/WFC3/F390W images, we select a blank-sky region in proximity to SDSS J0252+0039 and perform a power-spectrum analysis of the random surface-brightness fluctuations for different values of the drizzling parameters \textit{final pixscale} and \textit{final pixfrac}. We decrease the \textit{final pixscale} gradually, from the original value of 0.04 arcsec to 0.033, 0.025 and, finally, 0.02 arcsec. The \textit{final pixfrac} can be in principle varied between 0 (equivalent to sampling with a delta function) and 1 (drop size equal to the original pixel size), but we follow the recommended practice and set the drop size such that it is in each case slightly larger than the output pixels. Fig.~\ref{fig:pixscale_pixfrac_UV} presents the resulting azimuthally-averaged power spectra, in comparison to the default drizzling configuration. 

As can be seen from Fig.~\ref{fig:pixscale_pixfrac_UV}, lowering the \textit{final pixscale} (i.e. decreasing the output pixel size) and the \textit{final pixfrac} (i.e. shrinking the input pixels) when combining the dithered exposures does not allow us to substantially reduce the noise correlations i.e. flatten the noise power spectrum. Moreover, it has a significant effect on the measured power spectrum only in the highest-$k$ bins in which the residual surface-brightness fluctuations revealed in the modelled lens systems have the lowest amplitudes and are very close to the noise level. We stress that any choice of the \textit{final pixscale} and \textit{final pixfrac} values allows a valid analysis only if applied to both the science image and the blank-sky cutouts which are used for the estimation of the noise power spectrum. Taking into account that the choice of a lower output pixel scale, while maintaining the same field of view, would substantially increase the number of pixels in the analysed images and, thus, the computational effort of our study (especially the lens modelling), we conclude that the default configuration of the drizzling procedure (i.e. both the input and output pixel size equal to the original pixel size) is a suitable choice for our analysis.

\subsubsection{Effect of charge-transfer inefficiency}
\label{Section:CTI}

Due to a gradual degradation process of the HST/WFC3/UVIS-CCDs, the analysed images are additionally affected by the charge-transfer inefficiency \citep[see e.g.][]{Baggett_CTE}. This is caused by the radiation damage in space, which leads to defects in the silicon lattice of the CCDs and the formation of spurious trails in the observed images \citep{CTI}. 

\begin{figure}
\centering
\includegraphics[width=\columnwidth]{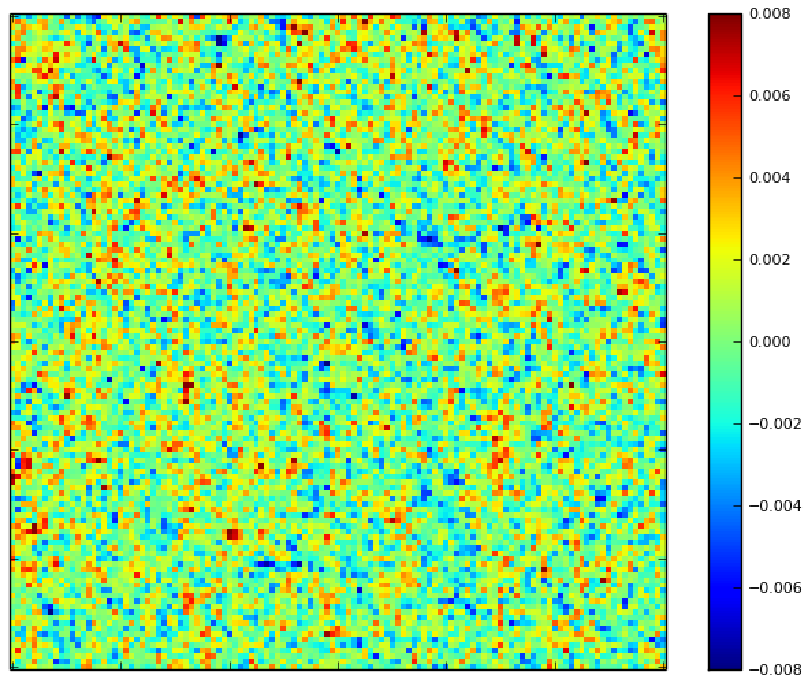}
\includegraphics[width=\columnwidth]{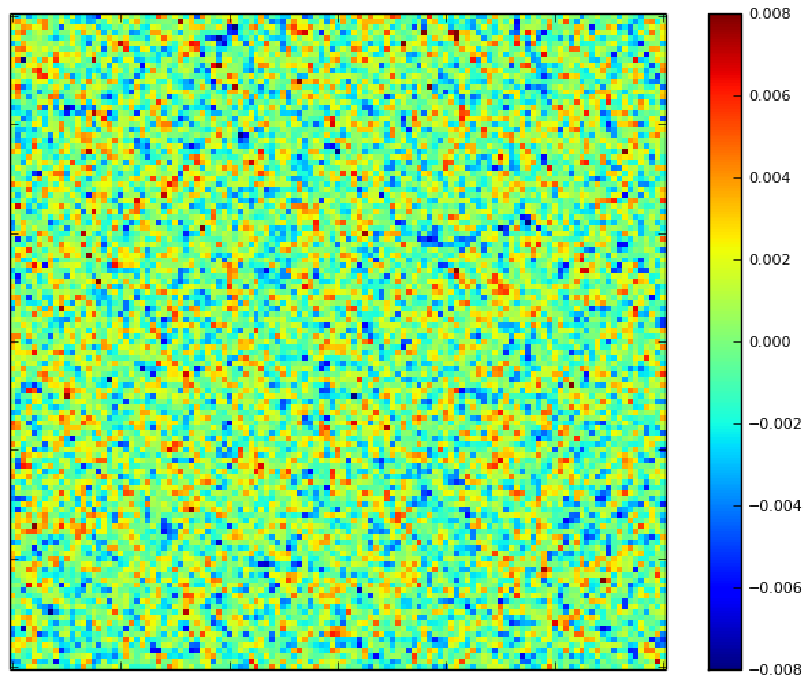}
\caption{The effect of the charge-transfer-efficiency (CTE) correction: a drizzled HST/WFC3/F390W blank-sky cutout located in proximity to SDSS J0252+0039 before \textit{(top panel)} vs. after the CTE correction \textit{(bottom panel)}. The correction allows us to significantly reduce the level of random surface-brightness fluctuations on large spatial scales, which becomes more apparent in Fourier space, see Fig.~\ref{fig:CTE_corr_PS}.}
\label{fig:CTE_corr_FITS}
\end{figure}

\begin{figure}
 \includegraphics[width=\columnwidth]{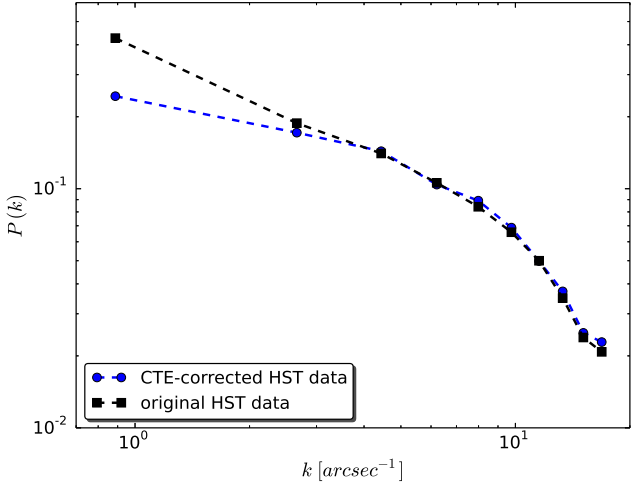}
 \caption{The effect of the charge-transfer-efficiency (CTE) correction on the azimuthally-averaged power spectrum of random surface-brightness fluctuations in a selected HST/WFC3/F390W blank-sky cutout located in proximity to SDSS J0252+0039 (corresponding to Fig.~\ref{fig:CTE_corr_FITS}).}
 \label{fig:CTE_corr_PS}
\end{figure}

In order to investigate the impact of this issue on the noise properties in our imaging, we perform the drizzling procedure of a selected blank-sky cutout located in vicinity to SDSS J0252+0039 using the charge-transfer-efficiency (CTE) corrected flat-field-calibrated exposures (flc.fits files) and compare the resulting image to the corresponding image based on the default flt.fits files. A careful visual inspection of these two images, presented in Fig.~\ref{fig:CTE_corr_FITS}, leads to the conclusion that the CTE correction allows us to reduce the level of random surface-brightness fluctuations on the largest spatial scales (smallest $k$-values). This effect becomes even more apparent in Fourier space. As can be seen from Fig.~\ref{fig:CTE_corr_PS}, the CTE correction results in a significant reduction of the noise variance on the largest considered spatial scales (more specifically, the power for $k_{\rm{min}}= $ 0.88 $\mathrm{arcsec^{-1}}$ corresponding to the spatial length scale $\lambda=1.13$ arcsec 
 or $\sim25$ pixels is roughly 40 per cent lower after the CTE correction). However, it does not affect the measured power spectrum on smaller spatial scales (corresponding to higher $k$-values).

While we recommend the use of CTE-corrected flat-field-calibrated exposures (flc.fits files) in future research, due to a substantial variance reduction of the sky-background fluctuations on the largest spatial scales, in the present paper and the accompanying Paper II we omit the CTE correction and proceed using the standard flt.fits files. We stress that the choice of either of the two options allows a valid analysis as long as both the science image and the blank-sky cutouts used to estimate the noise power spectrum are created in a consistent way (based either on the flt.fits or flc.fits files). Nevertheless, in order to test the impact of the CTE correction on our final results (i.e. exclusion probabilities of the matter-power-spectrum models presented in Paper II), we additionally compute the exclusion probabilities while excluding the largest considered spatial scales and find that this does not significantly affect the derived constraints. We plan to use CTE-corrected images and study this effect in more detail in the analysis of the Jackpot gravitational lens system SDSS J0946+1006, which will be presented in Paper~III of this series (Bayer et al., in prep).

\subsubsection{Estimation of the total noise power spectrum}
\label{Section:Noise_PS}

Here, we estimate the \emph{total} noise power spectrum in our HST-imaging which incorporates the combined effect of the random sky-background fluctuations, the noise correlations introduced in the drizzling procedure as well as the flux-dependent photon-shot noise from the observed lens system. To this end, we generate a set of \emph{scaled} sky-background cutouts, located in proximity to each analysed lens system, which can be seen as mock realisations of the total noise in the corresponding science image.

To create the scaled sky-background cutouts, we make use of the set of drizzled blank-sky cutouts generated in Section~\ref{Section:Noise_correlation_UV}. These can be treated (after subtracting the mean value) as realizations of a Gaussian random field with the expected value equal to zero and a constant standard deviation over the entire field-of-view, which we approximate by the standard deviation of the measured flux values. We first divide the blank-sky cutouts by this standard deviation to convert them to the standard normal distribution (i.e. the expected value equal to zero and the standard deviation equal to one) and, subsequently, multiply them by the respective noise-sigma map of the science image (see Section~\ref{Section:noise_sigma}) to incorporate the flux-dependent photon-shot noise from the observed lens system. The average power spectrum measured in the set of these scaled sky-background cutouts, located in proximity to the respective lens system, constitutes our best estimate of the total noise power spectrum in the observed science image. 

For a proper comparison with the measured power spectrum of the residual surface-brightness fluctuations in the lensed images (see Section~\ref{Section:smooth_lens_PS} and Fig.~\ref{fig:HSTn1vsNoise}), we perform the power-spectrum analysis of the (original and scaled) sky-background cutouts following the same methodology. In particular, before computing the azimuthally-averaged power spectrum as specified in Section~\ref{Section:PS_measurement}, we consistently overlay each sky-background cutout with the same mask outlining the lensed images as used in the analysis of the respective science image and set the remaining pixel values to zero. 

As an example, Fig.~\ref{fig:HSTn1vsNoise} shows the mean power spectrum of the sky-background fluctuations measured in the sample of twenty masked blank-sky regions in proximity to the lens system SDSS J0252+0039 and the total noise power spectrum estimated based on the corresponding scaled blank-sky cutouts, as discussed above. We stress that the deviation between the noise power spectra shown in Figs.~\ref{fig:HSTn1vsNoise} and~\ref{fig:correlated_empty_sky_UV} is due to the difference in the applied window function; that is, in Fig.~\ref{fig:HSTn1vsNoise} the power spectrum is computed based on masked cutouts for a consistent comparison with the residual power spectrum measured within the mask, while in Fig.~\ref{fig:correlated_empty_sky_UV} the cutouts are unmasked to investigate the pure effect of drizzling.

\subsection{Power spectrum of surface-brightness anomalies due to small-scale mass structures in the lens galaxy}
\label{Section:PS_SB_anomalies}

As a final step of our methodology, we perform a noise correction of the residual surface-brightness fluctuations and infer a (conservative) upper-limit constraint on the power spectrum of surface-brightness anomalies induced in the lensed images by the hypothetical small-scale mass structures in the lens galaxy. 

As thoroughly discussed in Section~\ref{Section:smooth_lens_PS}, we obtain our best estimate for the power spectrum of the residual surface-brightness fluctuations in the lensed images after the subtraction of the best-fitting smooth lens model inferred with a lower resolution (corresponding to $n=3$) and using the original tight mask (blue line in Fig.~\ref{fig:HSTn1vsNoise}). This choice allows us to prevent the overfitting problem as well as mitigate the degeneracy between the surface-brightness anomalies due to mass structures in the lens galaxy and the intrinsic surface-brightness fluctuations in the source galaxy itself. As can be seen from Fig.~\ref{fig:HSTn1vsNoise} for the lens system SDSS J0252+0039, the measured residual power spectrum exceeds in this case the noise power spectrum in the five lowest-$k$ bins (largest considered spatial scales) ranging from $k_{\rm{min}}= $ 0.88 to $k_{\rm{max}}= $ 7.95 $\mathrm{arcsec^{-1}}$, while it is at the noise level for all higher-$k$ modes.

\begin{figure}
\centering
\includegraphics[width=\columnwidth]{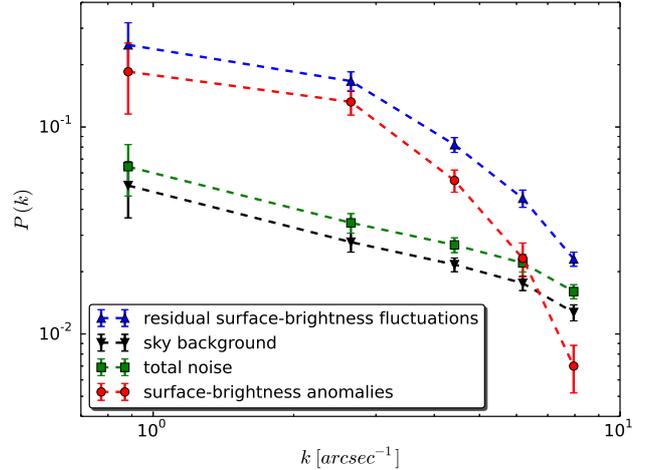}
\caption{Upper-limit constraints on the power spectrum of surface-brightness anomalies due to small-scale mass structures in the lens galaxy SDSS J0252+0039: the power spectrum of the residual surface-brightness fluctuations after the smooth lens modelling with $n=3$ \textit{(blue line)}, the mean power spectrum of the sky background in a sample of twenty blank-sky regions located in proximity to the lens \textit{(black line)}, the estimated total noise power spectrum including the photon-shot noise \textit{(green line)}, and the noise-corrected power spectrum of the residual surface-brightness fluctuations, constituting our upper limit on the power spectrum of surface-brightness anomalies due to small-scale mass structures in the lens galaxy \textit{(red line)}.}
\label{fig:power_spectra_UV}
\end{figure}

Assuming that the perturbations due to mass structure and the observational noise are statistically independent, we consider the corresponding power spectra to be additive. Under this assumption, we can simply subtract the estimated total noise power spectrum from the residual power spectrum. The procedure is illustrated in Fig.~\ref{fig:power_spectra_UV} for the lens system SDSS J0252+0039, where the difference of these two power spectra corresponds to the red line (shown only in the range of $k$-modes for which the residual exceeds the noise level). This noise-corrected power spectrum of the residual surface-brightness fluctuations in the lensed images constitutes our upper-limit constraint on the power spectrum of surface-brightness anomalies due to small-scale mass structures in the lens galaxy and is the final outcome of the methodology introduced in this paper. In the companion Paper II, we intend to extend our methodology and relate this measurement to the statistical properties of the underlying small-scale mass structures (more specifically, the \emph{sub-galactic matter power spectrum}) in the massive elliptical lens galaxy.

\section{Performance test}
\label{Section:Performance_test}

Finally, in this section, we test the performance of the introduced methodology in recovering the true power spectrum of mock surface-brightness anomalies from a simulated image mimicking real HST/WFC3/F390W-observations of the lens system SDSS J0252+0039, in which the lens galaxy is perturbed by small-scale sub-galactic mass structures with known statistical properties. 

We model these hypothetical small-scale mass structures in the lens galaxy as a realization of Gaussian-random-field (GRF) potential perturbations $\delta\psi_{\rm{GRF}}(\textbf{\textit{x}})$ superposed on a PEMD-plus-external-shear smooth lensing potential. Following \cite{Saikat}, we assume $\delta\psi_{\rm{GRF}}(\textbf{\textit{x}})$ to be fully characterised by a power-law power spectrum:
\begin{equation}
P_{\delta\psi}(k) = A \times k^{-\beta}
\label{eq:power_law_PS}
\end{equation}
with two free parameters, i.e. the variance of the potential perturbations $\sigma^2_{\delta\psi}$ and the power-law slope $\beta$. The power spectrum obeys the following normalization condition:
\begin{equation}
A \Big( \sigma^2_{\delta\psi}, \beta, L \Big) = \frac{L^2 \sigma^2_{\delta\psi}}{\sum_{k_{x}}\sum_{k_{y}} \Big( \sqrt{k_{x}^2+k_{y}^2} \Big) ^{-\beta}}, 
\label{eq:A}
\end{equation}
where the wavenumbers $k_{x}$, $k_{y}$ correspond to the reciprocal wavelength of the associated harmonic waves $e^{-2 \pi i \textit{k} \cdot \textit{x}}$ propagating in the $x$ and $y$ direction in the Fourier representation of the GRF and $L$ indicates the side length of the analysed image measured in arcsec, see Paper II for a more thorough discussion of the applied formalism. 

To simulate the effect of such potential perturbations on the lensed images of SDSS J0252+0039, we add a realization of  $\delta\psi_{\rm{GRF}}(\textbf{\textit{x}})$ to the best-fitting PEMD-plus-external-shear smooth lensing potential, inferred for this system in Section \ref{Section:Lens_Model_UV_sample}, and repeat the lensing operation of the reconstructed source galaxy. We set $\sigma^2_{\delta\psi} = 4 \times 10^{-4}$ and $\beta = 4$, such that the power spectrum of the induced mock surface-brightness anomalies resembles the residual power spectrum revealed in the real system. To account for the observational effects, we convolve this simulated image with the Tiny-Tim PSF of the HST/WFC3/F390W-optics and add a realistic noise realization. For simplicity, this noise realization is generated based solely on the noise-sigma map (see Section~\ref{Section:noise_sigma}) and, thus, does not reflect the noise correlations found in the drizzled images (see Section~\ref{Section:Noise_correlation_UV}). This simplifying assumption is justified by a low level of noise compared to the surface-brightness anomalies induced by the small-scale mass structures. 

We perform smooth lens modelling of this simulated image using the same methodology that was applied to the real observed data, within the same mask outlining the lensed images. The modelling is carried out without re-optimising for the parameter values of the best-fitting smooth lensing potential. By doing so, we assume that the parametric lensing potential can be reconstructed accurately and focus instead on investigating the degeneracy between the anomalies caused by the small-scale mass structures in the lens galaxy and the intrinsic surface-brightness fluctuations in the source galaxy itself. We will test the validity of this assumption in Paper III of this series (Bayer et al., in prep). In the current modelling procedure, we apply an adaptive source-grid regularisation and varying levels of the source-grid resolution, i.e. the number of pixels cast back from the lens plane to the source plane, corresponding to $n=1,2,3,4$ (see Section \ref{Section:Lens_Model_UV_sample}). In each case, we determine the resulting power spectrum of the residual surface-brightness fluctuations remaining in the lensed images after the subtraction of the best-fitting smooth lens model and compare it with the known true power spectrum of the imposed surface-brightness anomalies.

\begin{figure}
 \includegraphics[width=\columnwidth]{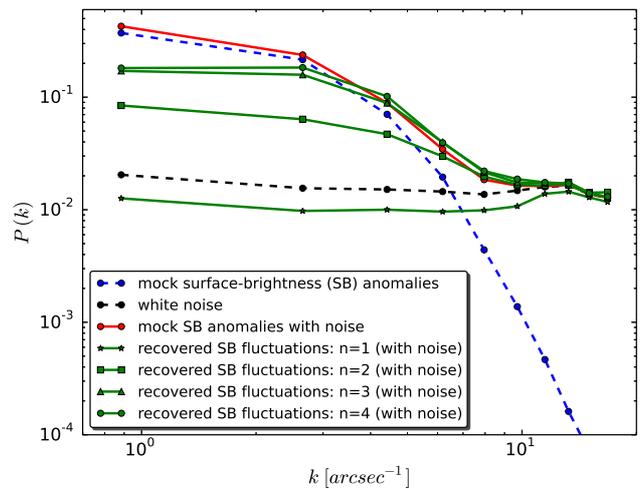}
 \caption{Performance test of the introduced methodology in recovering the power spectrum of mock surface-brightness anomalies induced in the lensed images of SDSS J0252+0039 by a realization of Gaussian random field potential fluctuations $\delta\psi_{\rm{GRF}}(\textbf{\textit{x}})$ with known statistical properties. The plot depicts the true azimuthally-averaged power spectrum of the induced surface-brightness anomalies \textit{(blue line)}, the power spectrum of the overlaid white-noise realization \textit{(black line)}, the power spectrum of these two effects combined (i.e. the power spectrum to be recovered, \textit{red line}) and the power spectra of the residual surface-brightness fluctuations actually recovered from the perturbed lensed images after the subtraction of the best-fitting smooth lens model obtained with varying source-grid resolution ($n=1,2,3,4$; \textit{green lines from bottom to top}).}
 \label{fig:mock_lens_test_UV}
\end{figure}

The results of this performance test are presented in Fig.~\ref{fig:mock_lens_test_UV}. From this figure, it can be seen that the power spectrum of the residual surface-brightness fluctuations lies significantly below the noise level when the modelling is performed with the highest source-grid resolution ($n=1$; i.e. each pixel is cast back from the lens plane to the source plane). As in the analysis of the real system, this overfitting can be explained by the absorption of the induced surface-brightness anomalies, and partially even the observational noise, in the source structure. However, this degeneracy can be alleviated by lowering the resolution (i.e. choosing higher $n$-values) of the adaptive source grid, which leads to a better agreement between the reconstructed and the true residual power spectrum. A comparison of the power spectra corresponding to $n=3$ and $n=4$ suggests that convergence is reached for $n=3$ and lowering the source-grid resolution even further does not allow us to thoroughly suppress this degeneracy (but would lead to a considerably less accurate source reconstruction). The absorption of the potential perturbations into the source structure persists on the smallest considered $k$-scales. 

All in all, based on the results of this performance test with a realistic mock lens, we conclude that our approach allows us to recover the true power spectrum of mock surface-brightness anomalies when the smooth lens modelling is performed with $n=3$ as the most suitable source-gird resolution for the given data quality and the choice of a relatively narrow mask.

\section{Conclusions and outlook}
\label{Section:Conclusions_UV_sample}

In this paper, the first in a series, we have introduced and tested a novel methodology to reliably measure the power spectrum of surface-brightness anomalies in extended lensed images of galaxy-galaxy strong gravitational lens systems. To illustrate our approach, we have applied it to a SLACS sub-sample observed with HST/WFC3 in the ultra-violet and discussed the modelling challenges. Finally, as a proof of concept, we have demonstrated the feasibility of the introduced methodology by recovering the true power spectrum of mock surface-brightness anomalies from simulated lensed images mimicking real HST/WFC3/F390W-observations of the lens system SDSS J0252+0039. 

One of the main challenges in the power-spectrum measurement turned out to be the degeneracy between the surface-brightness anomalies due to the presence of mass structures in the lens galaxy and the intrinsic surface-brightness fluctuations in the source galaxy itself. While this degeneracy is less problematic in the case of the direct detection of individual subhaloes with masses above the detection limit, as in \cite{Vegetti2014}, this issue requires a more careful consideration in the power-spectrum approach. Our test on simulated lensed images has shown that the degeneracy can be alleviated by performing the smooth lens modelling with a lower source-gird resolution to prevent overfitting. 

In the companion Paper II, our main objective is to extend the introduced methodology such that the estimated power spectrum of surface-brightness anomalies in the lensed images can be traced back to the statistical properties of the underlying small-scale mass structures in the lens galaxy. With this goal in mind, we carry out a systematic study of mock surface-brightness anomalies induced by Gaussian-random-field potential perturbations with varying statistical properties and compare the results to the real measurement performed in the present paper. For a pilot application of the extended methodology, we choose one of the lens systems from the investigated SLACS sub-sample, SDSS J0252+0039, due to its simple geometry and a high signal-to-noise ratio of the lensed images. As a final result of the combined analysis, we infer the first observational constraints on the matter power spectrum in a massive elliptical (lens) galaxy.

\section*{Acknowledgements}

The authors would like to thank the anonymous reviewer for his/her constructive and valuable comments on this work. We also thank Georgios Vernardos for his suggestions, many of which were very helpful. This study is based on observations made with the NASA/ESA \textit{Hubble Space Telescope}, obtained from the data archive at the Space Telescope Science Institute. Support for this work was provided by a VICI grant (project number 614.001.206) from the Netherlands Organization for Scientific Research (NWO) and by a NASA grant (HST-GO-12898) from the Space Telescope Science Institute. STScI is operated by the Association of Universities for Research in Astronomy, Inc. under NASA contract NAS 5-26555. DB acknowledges support by the Australian Research Council Centre of Excellence for All Sky Astrophysics in 3 Dimensions (ASTRO 3D), through project number CE170100013. TT acknowledges support by the Packard Foundation through a Packard Research Fellowship. CDF acknowledges support from the NSF under grant AST-1715611.

\section*{Data availability}
The images and mock data analysed in this work are available from the corresponding author upon reasonable request. The raw HST images are publicly available in the Mikulski Archive for Space Telescopes (MAST).



\bibliographystyle{mnras}




\appendix


\bsp	
\label{lastpage}
\end{document}